\documentclass[twocolumn]{aastex62}

\usepackage{ulem,xcolor}
\usepackage{epsf}
\usepackage{graphicx}
\usepackage{natbib}
\usepackage{url}
\usepackage{mathrsfs}
\usepackage{amsmath}
\usepackage{verbatim}

\newcommand{\Msun}{\mbox{\,$M_{\odot}$}}
\newcommand{\Lsun}{\mbox{\,$L_{\odot}$}}

\newcommand{\degree}{$^\circ$}

\graphicspath{{./}{figures/}}

\shorttitle{LUVIT I. Survey Overview}
\shortauthors{Gilbert et al.}

\newcommand{\marthago}{GO-16162}
\newcommand{\karriego}{GO-15275}
\newcommand{\ngals}{22}
\newcommand{\npointings}{23}

\newcommand{\distanceofmostdistantgal}{$3.8$~Mpc}

\newcommand{\astmagdiffthreshold}{0.75}

\newcommand{\CLmagbin}{0.2}

\newcommand{\spdclmagbin}{0.3}

\newcommand{\biasmagbin}{0.4}
\newcommand{\minstarsbiasmagbin}{25}

\newcommand{\spdbiasmagbin}{0.4}

\newcommand{\magbinphotunc}{0.2}
\newcommand{\magbinphotuncbincenter}{0.05}
\newcommand{\minstarsphotuncmagbin}{25}
\newcommand{\minastsphotuncmagbin}{100}

\begin{document}
\bibliographystyle{aasjournal}
\title{The Local Ultraviolet to Infrared Treasury I. Survey Overview of the Broadband Imaging}

\correspondingauthor{Karoline M. Gilbert}
\email{kgilbert@stsci.edu}

\author[0000-0003-0394-8377]{Karoline M. Gilbert}
\affiliation{Space Telescope Science Institute, 3700 San Martin Dr., Baltimore, MD 21218, USA}
\affiliation{The William H. Miller III Department of Physics \& Astronomy, Bloomberg Center for Physics and Astronomy, Johns Hopkins University, 3400 N. Charles Street, Baltimore, MD 21218, USA}

\author[0000-0003-1680-1884]{Yumi Choi}
\affiliation{NSF National Optical-Infrared Astronomy Research Laboratory, 950 North Cherry Avenue, Tucson, AZ 85719, USA}

\author[0000-0003-4850-9589]{Martha L. Boyer}
\affiliation{Space Telescope Science Institute, 3700 San Martin Dr., Baltimore, MD 21218, USA}

\author[0000-0002-7502-0597]{Benjamin F. Williams}
\affiliation{Department of Astronomy, University of Washington, Box 351580, Seattle, WA 98195, USA}

\author[0000-0002-6442-6030]{Daniel R. Weisz}
\affiliation{Department of Astronomy, University of California, Berkeley, Berkeley, CA, 94720, USA}

\author[0000-0002-5564-9873]{Eric F.\ Bell}
\affiliation{Department of Astronomy, University of Michigan, 1085 S.\ University Ave., Ann Arbor MI 48109, USA}

\author[0000-0002-1264-2006]{Julianne J.\ Dalcanton}
\affiliation{Center for Computational Astrophysics, Flatiron Institute, 162 Fifth Avenue, New York, NY 10010, USA}
\affiliation{Department of Astronomy, University of Washington, Box 351580, Seattle, WA 98195, USA}

\author[0000-0001-5538-2614]{Kristen B.~W.\ McQuinn}
\affiliation{Department of Physics and Astronomy, Rutgers, The State University of New Jersey, 136 Frelinghuysen Rd, Piscataway, NJ 08854, USA}
\affiliation{Space Telescope Science Institute, 3700 San Martin Dr., Baltimore, MD 21218, USA}

\author[0000-0003-0605-8732]{Evan D.\ Skillman}
\affiliation{Minnesota Institute for Astrophysics, University of Minnesota, 116 Church St.\ SE, Minneapolis, MN 55455, USA}

\author[0000-0002-6213-6988]{Guglielmo Costa}\affiliation{
Univ Lyon, Univ Lyon1, ENS de Lyon, CNRS, Centre de Recherche Astrophysique de Lyon UMR5574, F-69230 Saint-Genis-Laval, France.}

\author{Andrew E. Dolphin}
\affiliation{Raytheon, 1151 E. Hermans Road, Tucson, AZ 85756, USA} 
\affiliation{University of Arizona, Steward Observatory, 933 North Cherry Avenue, Tucson, AZ 85721, USA}

\author[0000-0001-9256-5516]{Morgan Fouesneau}
\affiliation{Max Plank Institute for Astronomy (MPIA), Königstuhl 17, 69117 Heidelberg, Germany.}

\author[0000-0002-6301-3269]{L\'eo Girardi}
\affiliation{INAF-Osservatorio Astronomico di Padova, Vicolo dell’Osservatorio 5, I-35122 Padova, Italy}

\author[0000-0002-8937-3844]{Steven R. Goldman}  
\affiliation{Space Telescope Science Institute, 3700 San Martin Dr., Baltimore, MD 21218, USA}

\author[0000-0001-5340-6774]{Karl D.\ Gordon}
\affiliation{Space Telescope Science Institute, 3700 San Martin Dr., Baltimore, MD 21218, USA}
  
\author[0000-0001-8867-4234]{Puragra Guhathakurta}
\affiliation{Department of Astronomy \& Astrophysics, University of California Santa Cruz, 1156 High Street, Santa Cruz, CA 95064, USA}

\author[0000-0003-3747-1394]{Maude Gull}
\affiliation{Department of Astronomy, University of California, Berkeley, Berkeley, CA, 94720, USA}

\author[0000-0001-8918-1597]{Lea Hagen}
\affiliation{Space Telescope Science Institute, 3700 San Martin Dr., Baltimore, MD 21218, USA} 
\affiliation{NXP Semiconductors, 10620 Treena Street, Suite 150/250, San Diego, CA 92131, USA}

\author[0009-0001-6590-4655]{Ky Huynh}
\affiliation{Space Telescope Science Institute, 3700 San Martin Dr., Baltimore, MD 21218, USA}

\author[0000-0003-0588-7360]{Christina W. Lindberg}
\affil{The William H. Miller III Department of Physics \& Astronomy, Bloomberg Center for Physics and Astronomy, Johns Hopkins University, 3400 N. Charles Street, Baltimore, MD 21218, USA}
\affiliation{Space Telescope Science Institute, 3700 San Martin Dr., Baltimore, MD 21218, USA}

\author[0000-0002-9137-0773]{Paola Marigo}
\affiliation{Department of Physics and Astronomy G. Galilei,
University of Padova, Vicolo dell'Osservatorio 3, I-35122, Padova, Italy}

\author[0000-0002-7743-8129]{Claire E. Murray}
\affiliation{Space Telescope Science Institute, 3700 San Martin Dr., Baltimore, MD 21218, USA}
\affiliation{The William H. Miller III Department of Physics \& Astronomy, Bloomberg Center for Physics and Astronomy, Johns Hopkins University, 3400 N. Charles Street, Baltimore, MD 21218, USA}

\author[0000-0002-9300-7409]{Giada Pastorelli}
\affiliation{Osservatorio Astronomico di Padova -- INAF, Vicolo dell'Osservatorio 5, I-35122 Padova, Italy}
\affiliation{Department of Physics and Astronomy G. Galilei, University of Padova, Vicolo dell’Osservatorio 3, I-35122, Padova, Italy}

\author[0000-0002-9912-6046]{Petia Yanchulova Merica-Jones}
\affil{Space Telescope Science Institute,
3700 San Martin Drive,
Baltimore, MD 21218, USA}

\begin{abstract}
The Local Ultraviolet to Infrared Treasury (LUVIT) is a Hubble Space Telescope program that combines newly acquired data in the near ultraviolet (NUV), optical, and near infrared (NIR) with archival optical and NIR imaging to produce multiband panchromatic resolved stellar catalogs for \npointings\ pointings in \ngals\
low-mass, star-forming galaxies ranging in distance from the outskirts of the Local Group to $\sim$\distanceofmostdistantgal. 
We describe the survey design,  detail the LUVIT broadband filter observations and the archival datasets included in the LUVIT reductions, and summarize the simultaneous multiband data reduction steps.  The spatial distributions and color-magnitude diagrams (CMDs) from the resulting stellar catalogs are presented for each target, from the NUV to the NIR. We demonstrate in which regions of the CMDs stars with NUV and optical, optical and NIR, and NUV through NIR detections reside. For each target, we use the results from artificial star tests to measure representative completeness, bias, and total photometric uncertainty as a function of magnitude in each broadband filter.  We also assess which LUVIT targets have significant spatial variation in the fraction of stars recovered at a given magnitude. 
The panchromatic LUVIT stellar catalogs will 
provide a rich legacy dataset for a host of resolved stellar population studies.

\end{abstract}


\section{Introduction}\label{sec:intro}

Nearby star-forming galaxies where individual stars can be resolved provide opportunities to perform accurate, quantitative tests of the time and energy scales associated with the cycle of gas and stars in galaxies, including 
star formation (SF) and feedback into the interstellar medium (ISM), stellar evolution, and
the baryonic processes driving galaxy formation \citep[e.g.,][]{weisz2008, dalcanton2009, gogarten2009b, gogarten2010, mcquinn2010a, weisz2011a, williams2011a, stilp2013}.  
The observed fluxes of individual stars in multiple wavelengths can be used to disentangle and make direct measurements of both 
stellar and dust parameters, and to derive maps of dust, SF, and its evolution over time on scales of tens to hundreds of parsecs \citep[e.g.,][]{dohm-palmer2002, skillman2003, weisz2008, mcquinn2012a, dalcanton2015, lewis2015, williams2017, Rubele2018, Cignoni2019, Ruiz-Lara2020, Lazzarini2023}.  Observations spanning from the near ultraviolet (NUV) through the near infrared (NIR) yield crucial observational constraints on the evolution and impacts of the massive young stellar populations and the evolved, dust-producing stellar populations in galaxies. 

NUV observations of resolved stars provide a sensitive probe of the youngest SF timescales, providing an independent estimate of the SF in galaxies over the timescales probed by common tracers of the recent star formation rate (SFR) \citep[e.g.,][]{Calzetti2015, Cignoni2019, Sacchi2019}.  They also enable inference of the far UV (FUV) flux produced by the youngest stars and the amount of UV flux obscured by dust \citep{gogarten2009a, johnson2013, simones2014, lewis2017}, providing a mechanism for estimating the high energy, ionizing flux deposited into the galactic ecosystem by young stars \citep[e.g.,][]{Leitherer1996, Crowther2010, Lopez2011,  bouret2012, choi2020}.  

In the optical and NIR, bright, evolved stars (e.g., thermally pulsing asymptotic giant branch (AGB) stars and red helium-burning (RHeB) stars) are a dominant source of the flux in star-forming galaxies \citep{melbourne2012,melbourne2013}, strongly influencing quantities derived from models in more distant galaxies \citep[e.g., the SFR, total stellar mass, and metallicity;][]{conroy2009,baldwin2018}.   
Observations of evolved stars across a wide range of galaxy masses and metallicities provide critical constraints on stellar evolutionary models, including the processes that drive mass loss, dredge up, convective overshoot, and rotation \citep{conroy2009, girardi2010, marigo2013, rosenfield2014, tang2014, choi2016,  marigo2017, costa2019, pastorelli+2019, pastorelli+2020, nguyen2022}.

Over the past three decades, the Hubble Space Telescope (HST) has imaged resolved stellar populations in the optical
throughout the Local Volume providing a rich and informative optical legacy, while imaging of resolved stars in the NUV and NIR has been limited to a narrower range of targets. Concerted efforts such as the ACS Nearby Galaxy Treasury 
(ANGST; PI Dalcanton, GO-10915, DD-11307;  \citealt{angst_doi,dalcanton2009}) and the ACS Nearby Galaxy: Reduce, Reuse, and Recycle 
(ANGRRR; PI Dalcanton, AR-10945; \citealt{angrrr_doi}) synthesized much of this optical imaging into a uniform and publicly accessible database.
The resolved stellar imaging legacy of HST in the optical has been augmented over the last decade and a half by NUV through 
NIR imaging of resolved stellar populations in the Local Group,
including the Panchromatic Hubble Andromeda Treasury in M31~\citep[PHAT; PI Dalcanton, GO-12055;][]{dalcanton2012phat}, the Panchromatic Hubble Andromeda Treasury: Triangulum Extended Region in M33 \citep[PHATTER; PI Dalcanton, GO-14610;][]{williams2021phatter}, the Hubble Tarantula Treasury Project in the LMC \citep[HTTP; PI Sabbi, GO-12939;][]{sabbi2013}, the Small Magellanic Cloud Investigation of Dust and Gas Evolution in the SMC \citep[SMIDGE; PI Sandstrom, GO-13659;][]{smidgepetia2017}, and Scylla in the LMC and SMC (PI Murray, PIDs 15891, 16235, and 16786; Murray et al., submitted).
NUV through NIR imaging has also been undertaken in more distant galaxies ($\sim 3.5$ to 12 Mpc) by
the Legacy Extragalactic Ultraviolet Survey
\citep[LEGUS; PI Calzetti, GO-13364;][]{Calzetti2015}.  

Together, the above surveys provide panchromatic resolved stellar imaging of the four most massive galaxies in the Local Group (M31, M33, LMC, and SMC), as well as galaxies covering a wide range in mass and metallicity, primarily at distances $>3.5$~Mpc (LEGUS).  
The above surveys primarily
trace higher mass, higher metallicity star-forming
environments than those found in the 
lower-mass, metal-poor galaxies which 
form the bulk of the number density of galaxies in the Universe and
typically have high specific SFRs.  

LUVIT (PIs K.~M.~Gilbert and M.~L.~Boyer, \karriego\ and \marthago, respectively) was designed to target the range of observational parameter space not covered by the above surveys. The aim of LUVIT is to obtain panchromatic measurements 
of the resolved stellar populations of a sample of 22 low-mass, low-metallicity, nearby star-forming galaxies, extending HST's resolved stellar populations legacy to include NUV through NIR observations of the lowest mass ($M_* \lesssim 10^{7.5} M_\odot$) and lowest metallicity ([M/H]\,$\lesssim$\,12\% solar) star-forming environments.  

The LUVIT targets were chosen to cover all known star-forming systems closer than $\sim 3.5$~Mpc (which minimizes stellar-crowding effects), with estimated stellar masses $M_* \lesssim 10^{7.5}$\Msun, for which ancillary HST-optical, H$\alpha$ and GALEX FUV imaging is available.
The resulting multiwavelength datasets will enable comparisons between the dust content derived from the NUV through NIR stellar catalogs and the values inferred from commonly used ISM tracers, as well as calibration of widely used, multiwavelength SFR estimators \citep[e.g.,][]{simones2014, mcquinn2015a, mcquinn2015b}. 

The \karriego\ NUV observations were optimized for studying the young massive star populations.  One of the primary analysis goals of \karriego\ is to constrain the recent star formation history (SFH) of LUVIT galaxies with fine time and spatial resolution via quantitative color-magnitude diagram fitting techniques.  Another primary analysis goal is to infer the stellar properties, via stellar spectral energy distribution (SED) fitting techniques, of the young stellar sources contributing the majority of the FUV and NUV flux.    
The \marthago\ NIR and supplementary optical observations were optimized for studying the evolved star populations in these metal-poor galaxies, 
with the goal of measuring the observables (luminosity functions, stellar counts, CMD morphology, stellar SEDs) needed to constrain metallicity dependencies on the processes that drive evolved star model uncertainties. 
LUVIT's new observations in the NUV to NIR supplement archival optical and NIR medium or broadband Hubble data, allowing simultaneous photometric measurements of detected sources across the full NUV to NIR regime.  

This paper provides an overview of the LUVIT survey design and data reduction strategies (including the resolved stellar photometric measurements and artificial star tests), presents broadband CMDs across the full NUV through NIR wavelength range, and assesses the data quality as a function of magnitude for all broadband data (including \karriego, \marthago, and archival datasets).  A future companion paper will present the NIR medium-band imaging and a public data release of the full photometric catalogs (M.~Boyer et al., in prep).

This paper is organized as follows. Section~\ref{sec:survey_design} describes the sample selection and observational strategy for the \karriego\ and \marthago\ observations, and the identification of archival datasets for inclusion in the photometric reductions.  
Section~\ref{sec:data_redux} describes the process used to obtain simultaneous panchromatic photometric measurements from the multicamera, multi-instrument Hubble imaging. 
Section~\ref{sec:data} presents the spatial distribution and CMDs of sources in the resulting stellar catalogs. Section~\ref{sec:data_redux_asts}
describes the artificial star tests that were used to determine the photometric quality of each LUVIT pointing and presents representative estimates of the completeness, bias, and photometric uncertainties of the stellar catalogs.  It also assesses which targets display a significant dependence of these quantities on the observed stellar density.  Section~\ref{sec:notes_select_targets} discusses special considerations or issues relevant to select targets. Section~\ref{sec:science_apps} describes select science applications enabled by the LUVIT photometric catalogs.  Section~\ref{sec:conclusions} summarizes the results.

\section{Survey Design}\label{sec:survey_design}
\subsection{Survey Sample}\label{sec:sample}

The 22 LUVIT NUV (\karriego) galaxies were chosen for their high legacy value.  They span  
a significant range of metallicity, gas fraction, SFR intensity, and environment.  Several criteria were implemented in order to restrict the sample from the $\gtrsim 100$ galaxies in the Local Volume.  Building off the foundational work of ANGST, as well as earlier HST optical campaigns \citep{tully2006,dalcanton2009}, the LUVIT sample was restricted to galaxies with distances less than 3.5~Mpc\footnote{Since the selection of the LUVIT sample, revised distance estimates for one of the targets, UGCA292, place it at a distance greater than the 3.5~Mpc cutoff.}, to mitigate the effects of stellar crowding. 
Furthermore, only galaxies with estimated stellar masses $\lesssim 10^{7.5}$\Msun\ were included in order to target the lowest mass, lowest metallicity star-forming systems in the local universe.  The sample was further restricted to galaxies with existing optical HST data in at least two broadband filters, further requiring the data be of sufficient depth to (i) constrain main sequence (MS) stars with masses $> 5$\Msun\ ($M_V\sim -0.5$), which are responsible for $\gtrsim 70\%$ and $\gtrsim 60\%$ of the FUV and NUV flux, respectively \citep{johnson2013}, and (ii) detect and measure red giant branch (RGB) stars $>2$ magnitudes fainter than the tip of the RGB.  These fainter RGB stars are needed to anchor older SFR estimates and to provide a dense network of dust extinction probes.  Finally, a galaxy was also required to have existing, rich ancillary datasets.  Each galaxy was required to have H$\alpha$ and H{\sc i}, many were observed by GALEX and Spitzer, and most have also been targeted by other major facilities across the electromagetic spectrum, e.g., {\emph{Herschel}}, {\emph{VLA}}, {\emph{CARMA}}, and {\emph{IRAM}}. 

Following the NUV observations, we obtained additional NIR and select optical imaging in \marthago\ for all but three of the most distant LUVIT NUV targets (DDO6, M81-DWARF-A, and UGC8760). The omission of these three galaxies substantially reduced the orbital request of the program with minimal impact to the science goals of the \marthago\ observations.  

\begin{figure*}
    \centering
    \includegraphics[width=\textwidth]{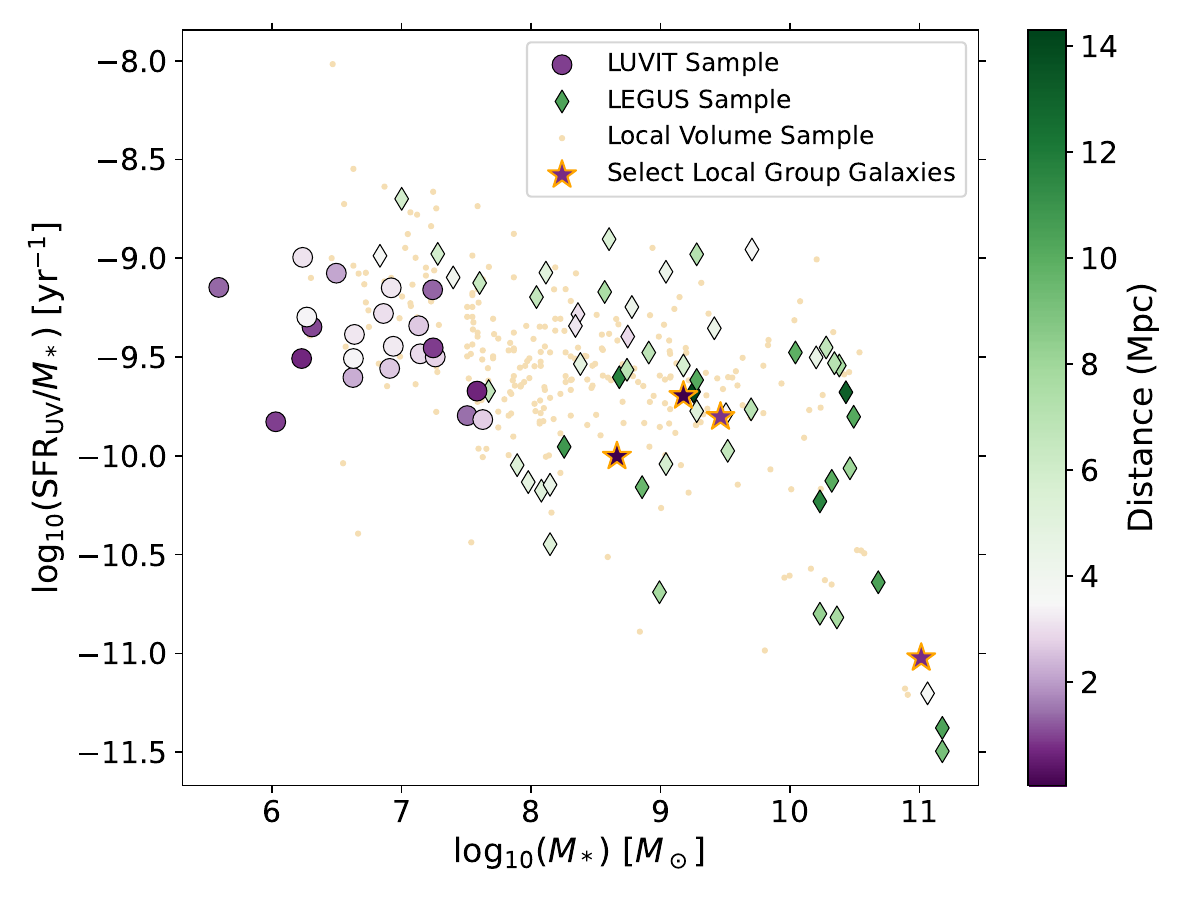}
    \caption{The specific SFR of star-forming galaxies in the local universe as a function of their stellar mass, drawn from the 11~Mpc sample of \citet{lee2009}.  The larger points represent the LEGUS (diamonds), LUVIT (circles), and select Local Group (M31, M33, LMC, and SMC; stars) galaxies and are color coded by distance, with the divergent point of the color map set at 3.5~Mpc, which is the approximate upper and lower limits of the LUVIT and LEGUS surveys, respectively.  The LUVIT survey is complementary to the LEGUS survey, with the LUVIT galaxies chosen to represent the lowest stellar mass, lowest metallicity star-forming galaxies in the local universe (Section~\ref{sec:sample}), with estimated stellar masses $M_* \lesssim 10^{7.5}$\Msun.  The resulting sample has gas-phase oxygen abundances of $12 + \rm{log(O/H)} < 8$ (Table~\ref{table:galaxysample}).
    }
    \label{fig:sample}
\end{figure*}

Implementing the above four criteria resulted in a sample that spans the remaining parameter space in the specific SFR\,--\,stellar mass plane which was not covered by LEGUS or existing Local Group surveys.  Figure~\ref{fig:sample} compares the specific star formation rate (SSFR) as a function of stellar mass ($M_*$) of the LUVIT and LEGUS samples against a broader sample of star-forming galaxies within the Local Volume.   The Local Volume sample is that of \citet{lee2009}, and includes all star-forming galaxies within 11~Mpc of the Milky Way.
The  SFR and $M_*$ values for the LEGUS galaxy sample were taken from \citet[][Table~1]{Calzetti2015}  for the galaxies observed by LEGUS, and taken or derived from \citet[][Table~1]{lee2009} for the galaxies for which LEGUS used archival datasets.  In both the 11~Mpc Local Volume sample and the LEGUS sample, the SFR  was calculated from the \textit{GALEX} FUV flux after correcting for dust attenuation.  Following \citeauthor{lee2009}\ and \citet{Calzetti2015}, the stellar mass estimates are calculated from the absolute $B$-band magnitude of each galaxy, using a morphology-color relationship determined by \citet{bothwell2009} and mass-to-light ratio models by \citet{bell2001}.   The values for LUVIT were similarly derived from \citeauthor{lee2009}\ where possible, or from the Updated Nearby Galaxy Catalog \citep[UNGC;][]{karachentsev2013ungc,karachentsev2013sfprop}. 
The stellar masses for the  Local Group galaxies (M33, M31, LMC, and SMC) are taken from \citet[M31]{sick2015} and \citet{mcconnachie2012}, while the SFRs are taken from the UNGC \citep{karachentsev2013sfprop}.\footnote{For three galaxies (UGC8508, LMC, and SMC), a FUV SFR was not available in any of the above references and the H$\alpha$-based SFR from \citet{karachentsev2013sfprop} is plotted instead.}  The LUVIT and LEGUS galaxies are color coded by distance, with the color map diverging at 3.5~Mpc, the approximate upper and lower limit of the LUVIT and LEGUS surveys, respectively.   As described above, the LUVIT galaxies were chosen as representatives of the lowest stellar mass, star-forming galaxies in the local universe.

Table~\ref{table:galaxysample} lists basic properties of the LUVIT galaxies, as well as commonly used alternate names, the filters obtained as part of LUVIT (from both \karriego\ and \marthago\ programs), and the archival filters included in the LUVIT reductions.  As noted, many of the galaxy measurements are taken from the Catalog and Atlas of the LV galaxies (LVG; http://www.sao.ru/lv/lvgdb/), which is the online portal of the Updated Nearby Galaxy Catalog \citep{karachentsev2013ungc}.  Interested readers are referred to either resource for detailed descriptions of the measurements or derivations of each quantity.

\subsection{Observational Strategy}\label{sec:obs_strategy}

\subsubsection{NUV Observations}\label{sec:obs_strategy_nuv}

The pointing and orientation of the \karriego\ UV observations were chosen with the aim of maximizing coverage of the star-forming regions of the galaxy, as well as overlap with existing broad band archival observations.  
For the majority of targets, a single WFC3/UVIS field of view covers most, or all, of the current SF (Figure~\ref{fig:footprints}).  Exceptions are primarily galaxies within or at the edge of the Local Group (e.g., IC1613, Sextans A, WLM). Two pointings were obtained in WLM, since a second pointing substantially increased coverage of the star-forming regions while taking maximum advantage of the existing deep archival optical observations.  

The
observations were restricted to a limited range of orientations rather than a single value; an acceptable range was determined for each target that enabled reasonable scheduling windows while maintaining the largest possible overlap with both the regions of active SF and the spatial extent of the existing archival observations.  

However, one pointing in Wolf–Lundmark–Mellote (WLM; WLM-POS1) was intentionally partially offset from the existing deep archival optical imaging to overlap ALMA CO detections.  Since WLM is one of the most metal-poor galaxies in which CO has been detected \citep{elmegreen2013,rubio2015}, we judged the importance of observing the stellar populations in the region of the CO detections to outweigh the goal of maximizing overlap with the archival optical imaging.  We subsequently obtained F475W observations with WFC3/UVIS which fully overlap the WLM-POS1 UV observations in \marthago.\footnote{The \karriego\ and \marthago\ observations also cover the confirmed CO detections in Sextans~B, which is currently the most metal-poor galaxy in which CO has been detected \citep{shi2016,shi2020}. Contemporaneously obtained observations from GO-16104 (P.I. J.~Roman-Duval), also included in our reductions, cover a marginal CO detection in Sextans~A \citep[][Section~\ref{sec:archival_datasets}]{shi2015}.}

The filters F275W and F336W were chosen to balance the depth of the observations and the overall orbital request, and to minimize the impact of the 2175\AA\ extinction feature.  We observed all galaxies at distances $<2.5$~Mpc for two orbits, and galaxies at distances $>2.5$~Mpc for three orbits, with the total exposure time split approximately equally between the F275W and F336W filters.  

The exposure specifications within and amongst orbits were chosen to maximize the total exposure time, cover the gap between the two WFC3/UVIS chips, and maintain approximately equal exposure times in the F275W and F336W filters, all while obtaining the best possible sampling of the point spread function. 
For the first orbit on each target, we performed a four-point dither pattern in F336W.  We used the POS TARG functionality in APT to offset each dither position from the nominal RA and Dec of the target position, in a pattern consisting of two small subpixel dithers, followed by a larger third dither to cover the chip gap, and finally a fourth small subpixel dither. The POS TARG (X, Y) positions, relative to the target RA and Dec, for these four dithers were ($-0.1287$, $-1.006$), ($-0.0297$, $-1.0586$), (0.0891, 0.9837), and (0.1881, 1.0497) arcseconds. The second orbit on each target repeated the same pattern of 4 pointings with F275W.  For targets with three orbits, the final orbit was split between F275W and F336W, with dithers between the two exposures in each filter large enough to provide coverage of the chip gap: the POS TARG positions defining the dithers were ($-0.2277$, $-0.95365$) and (0.2871, 1.1157) arcseconds.  To obtain the maximum possible exposure time, we
 used the Orbit Planner functionality in APT to determine exposure times for each dither position that resulted in as tightly ``packed'' an orbit as possible within the visibility window.  This resulted in two slightly shorter and two slightly longer exposures per orbit.  To help mitigate and control for charge transfer efficiency losses, we also implemented the minimum recommended flash setting in APT (a setting of 9 or 10 for our observations), which specifies the number of electrons per pixel to add to each image using the post-flash LED.  

\subsubsection{NIR and Supplemental Optical Observations}

The pointings of the \marthago\ optical and NIR observations were chosen to maximize overlap with the existing \karriego\ and archival broadband optical (and when available, broadband and medium-band NIR) data. Fourteen \marthago\ targets were observed in the WFC3/NIR F110W and F160W broadband filters, and 19 in the F127M, F139M, and F153M medium-band filters. The broadband NIR data were obtained for identification of RHeB stars, which are not easily separated from RGB and thermally pulsing AGB (TP-AGB) stars in the optical, and for sampling the peak bolometric flux of the RGB, RHeB, and TP-AGB stars.  The medium-band data were obtained to classify the chemical types (carbon-rich or oxygen-rich) of the TP-AGB stars using the NIR molecular features of TP-AGB stars, particularly the CN+C2 in carbon stars and the water features near 1.4~$\mu$m for oxygen-rich TP-AGB stars.  

Targets within 2.2~Mpc were observed with 1 orbit split approximately evenly between F110W and F160W, 
while more distant targets were observed with one orbit in each of F110W and F160W. 
The medium-band observations were observed in one orbit per target, with $\sim 800$\,--\,900~s exposures obtained in each filter, and will be described in detail in Boyer et al. (in prep).  

For one intended \marthago\ target, UGC8091, the planned wide-band IR observations were determined to be a duplication of observations being obtained in the same cycle by GO-16292 (PI Choi), and we therefore use the observations from GO-16292, 
bringing the total of LUVIT galaxies with new broadband NIR observations to fifteen.  The GO-16292 broadband NIR observations of UGC8091 obtained exposures at two dither positions in F110W and at eight dither positions in  F160W. 

The detailed exposure specifications of the \marthago\ WFC3/IR F110W and F160W observations varied depending on whether parallel observations with ACS/WFC (Section~\ref{sec:parallel_datasets}) were being obtained for that target, and if so, whether medium band filter observations were also being obtained, as that affected the structure of the parallel observations.  

The F110W and F160W exposures for targets without parallel observations were obtained using a four-point dither pattern, utilizing the default values of the \textsc{WFC3-IR-DITHER-BOX-MIN} pattern in order to sample the point spread function (a point spacing 0.572\arcsec, a line spacing of 0.365\arcsec, a pattern orientation of 18.528\degree, and the angle between sides set at 74.653\degree).  For targets in which the broadband NIR filters were observed in one orbit total (Leo~A, Antlia, and SagDIG), each filter was observed with the full four-point dither pattern and the orbit was evenly split between the two filters.  For targets in which the broadband NIR filters were observed over two orbits (UGC~7577, UGC~8651, UGC~8833, UGC~9128, UGC~9240), one orbit was dedicated to each of the broadband NIR filters, again observed using the full four-point dither pattern.  

For targets with associated parallels, the number of dither positions obtained in each filter and the length of the exposure times at each dither position were set to 
maximize the exposure time in both the primary NIR and parallel observations while minimizing overhead and managing data download constraints. For UGC6817, for which parallels were obtained and F110W and F160W were observed for one orbit each, exposures in both F110W and F160W were obtained at each dither position.  For the four targets with parallels and with one orbit split between the broadband filters and one orbit split between the medium-band filters (DDO210, Sextans~A, WLM-POS1, WLM-POS2), F160W exposures were obtained at all four dither positions while F110W exposures were obtained only at the third and fourth dither positions.  For Sextans~B, for which parallels were obtained and one orbit was split between the broadband filters but no medium-band observations were obtained, two exposures in F160W were obtained at dither position 1, two exposures in F110W were obtained at dither position 2, and one exposure in each of F160W and F110W were obtained at dither position 3.

Finally, an additional $\sim$100\,--\,200~s exposure was obtained at the first dither position for UGC6817, UGC7577, UGC8651, UGC8833, UGC9128, and UGC9240. All six of these targets were observed with one orbit for each observation in the F110W and F160W filters, regardless of the inclusion of parallels, and the additional short exposure maximized the total NIR exposure time given the available orbit visibility.   

\marthago\ also observed eleven galaxies with WFC3/UVIS in F475W, using one orbit per target.  In these galaxies, F555W or F606W were the shortest wavelength archival optical data with sufficient depth to meet the survey requirements. The addition of the F475W filter aids in identifying blue He-burning (BHeB) stars by increasing the color baseline, while also decreasing uncertainties in parameters derived via SED fitting by providing a significantly better constraint to the amplitude of the Balmer jump \citep{dalcanton2012phat}. 
The WFC3/UVIS F475W observations were obtained using a three point dither pattern, utilizing the default values of the \textsc{WFC3-UVIS-LINE-3PT} pattern in order to sample the point spread function (a point spacing of 0.135\arcsec\ and a pattern orientation of 46.84\degree).  The exposure times were chosen to fill the visibility window of the orbit. 

\subsubsection{Summary of LUVIT Observations}

\begin{figure}
    \centering
    \includegraphics[width=\columnwidth]{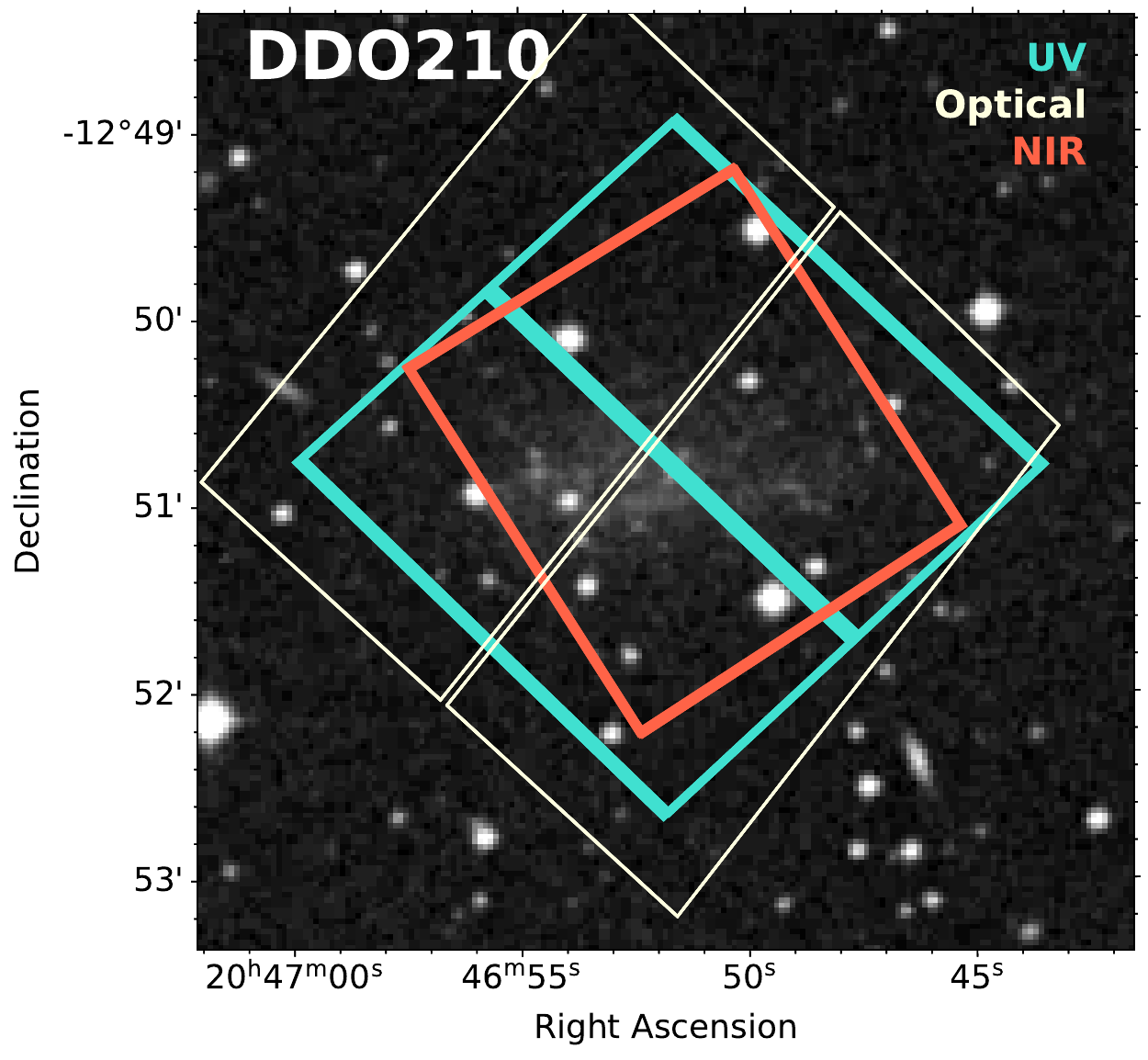}
    \caption{
    Footprints of the LUVIT GO and archival observations utilized in DDO210, overlaid on a Digitized Sky Survey image.  Observations obtained as part of \karriego\ or \marthago\ (Section~\ref{sec:obs_strategy}) are displayed with thick lines, while archival observations (Section~\ref{sec:archival_datasets}) are displayed with thin lines.  Turquoise footprints denote UV observations, light yellow footprints denote optical observations, and red footprints denote NIR observations. When the observations were taken with an instrument with more than one chip (all but the WFC3/IR camera), the footprint shows the location of the chip gap(s). The complete figure set (23 images) is available in the online journal and includes a figure for each LUVIT pointing.  
    }
    \label{fig:footprints}
\end{figure}

The combination of the \karriego, \marthago, and archival observations results in NUV through NIR imaging for 19 
of the 22 LUVIT galaxies (20 of the 23 LUVIT pointings).  
Figure~\ref{fig:footprints} shows, for each galaxy, the footprints of the WFC3/UVIS LUVIT NUV observations (\karriego), WFC3/IR, and WFC3/UVIS optical LUVIT observations (\marthago), and the archival optical and NIR data included in the LUVIT reductions (Section~\ref{sec:archival_datasets}).  The HST footprints are overlaid on Digitized Sky Survey (DSS1) images\footnote{Based on photographic data obtained using the UK Schmidt Telescope. The UK Schmidt Telescope was operated by the Royal Observatory Edinburgh, with funding from the UK Science and Engineering Research Council, until 1988 June, and thereafter by the Anglo-Australian Observatory. Original plate material is copyright © the Royal Observatory Edinburgh and the Anglo-Australian Observatory. The plates were processed into the present compressed digital form with their permission. The Digitized Sky Survey was produced at the Space Telescope Science Institute under US Government grant NAG W-2166.}. 

Tables~\ref{table:primarypointings} and \ref{table:martha_primarypointings} provide a basic summary of each of the primary \karriego\ WFC3/UVIS and \marthago\ WFC3/IR and WFC3/UVIS LUVIT pointings, respectively. 
These tables include both the Mikulski Archives for Space Telescopes (MAST) target name  
and the target name adopted in the primary LUVIT reductions. For all but three targets (which are not included in \marthago), this is the reduction that includes \karriego, \marthago, and archival data. 
For the three targets not included in \marthago\ (DDO6, M81-DWARF-A, and UGC8760), the pipeline name comes from initial reductions of \karriego\ and archival data (Section~\ref{app:stack_comparison}). 
Depending on context, we use either the galaxy name or the closely related pipeline name from the primary reductions for the identification of targets throughout the remainder of this paper.  

\subsection{Archival Datasets}\label{sec:archival_datasets}
  
Upon execution of the UV observations for each target, we searched the MAST archive to ensure we identified all relevant archival observations. In particular, we identified, for each 
\karriego\ pointing, the maximally overlapping archival imaging in broad optical or NIR bands, from any of Hubble's primary imaging cameras: WFPC2, ACS/WFC, WFC3/UVIS or WFC3/IR.  These datasets were subsequently included in our simultaneous, multiband reductions (Section~\ref{sec:data_redux}).  While data of sufficient depth in at least two optical broadband filters was a requirement for inclusion in the sample (Section~\ref{sec:sample}), a quarter of the sample galaxies also have archival WFC3 NIR imaging \citep[the majority from ANGST-SNAP;][]{dalcanton2012NIR}.  

In general, we did not include shallower, partially overlapping archival data when deeper, fully overlapping data existed in the same (or similar) bands. In some targets, deeper data with newer cameras (e.g., ACS/WFC or WFC3/UVIS) substantially overlapped significantly shallower data obtained with WFPC2: in these cases, we used only the ACS or WFC3/UVIS data\footnote{In addition to deeper photometry, the ACS and WFC3/UVIS instruments have fewer chips and associated chip gaps, and simpler footprints, than WFPC2, reducing the spatial areas with differing and complex completeness functions in the multiband stellar catalogs.}. 
However, for four targets, archival observations from different programs taken in the same filter and camera combinations were included.  The data from the multiple programs either increased the spatial area of the archival imaging in a given filter relative to the UV observations (Sextans~A, UGC7577) or substantially increased the exposure time of the archival imaging (NGC4163, SAGDIG). 

In addition, four targets had deep medium-band imaging in the archive (IC1613, SAGDIG, Sextans~A, and Sextans~B), which we also included in our multiband reductions.  The medium-band archival imaging was in F127M, F139M, and F153M with WFC3/IR for SAGDIG, Sextans~A, and Sextans~B (GO-14073, PI M.~Boyer) and in F390M with WFC3/UVIS for IC1613 (GO-12304, PI J.~Holtzman).  

For a small number of targets, existing archival optical imaging was extremely deep (DDO210, LEO-A, SAGDIG, and WLM-POS2). Including all the available archival exposures would have resulted in a depth well beyond that which is necessary for the goals of the LUVIT program, and would have needlessly increased the required computational resources for analyzing these targets. Therefore, in these cases we limited the archival imaging to the first full dither pattern\footnote{For Sag-DIG, the deep imaging (GO-12273; PI R.\ van der Marel) was in F814W only and taken at 3 primary positions, each with its own small dither pattern.  Since the deep observations overlapped previously existing F814W observations (GO-9820; PI Y.\ Momany) we included the full small dither pattern of the first of the primary positions, choosing better PSF sampling over increased chip gap coverage.} of the deep archival observations in each band, ensuring we recovered the best possible sampling of the point s
pread function and coverage of chip gaps.  This strategy still resulted in depths in the archival bands which exceed that of the typical target in the LUVIT survey. 

We included data obtained concurrent to and after the \karriego\ 
observations in one target.  Partially overlapping WFC3/UVIS observations in Sextans~A were obtained by GO-16104 (PI J.~Roman-Duval), in the F225W, F275W, F336W, F475W, and F814W filters.  These observations were designed to support the Hubble Director's Discretionary program ULYSSES, which obtained UV spectra of massive stars in the local universe \citep{ullysesRNAAS}.  The GO-16104 observations extend the UV and optical coverage of the star-forming region of Sextans~A, with the footprints of the observations including the marginal IRAM 30~m telescope CO detection by \citet{shi2015}. All the GO-16104 observations are at the same pointing; their position in Figure~\ref{fig:footprints} is represented by the archival UV observation footprint located below and partially overlapping with the LUVIT observation footprints.

The total exposure time in the archival bands varies widely, as did the science goals of the various GO programs. 
Table~\ref{table:archivalpointings} provides details of the archival observations used in the LUVIT reductions.       

\subsection{Parallel Datasets}\label{sec:parallel_datasets}
Both \karriego\ and \marthago\ obtained parallel imaging along with the primary observations.  While we do not present the results of the parallel imaging in this survey overview, for completeness we summarize the observational strategies used in obtaining the parallel exposures below.

Optical ACS/WFC images in F606W and F814W were obtained in parallel for all of the \karriego\ pointings. The parallels were obtained to enable the spatial extent to be determined and the ages of any extended stellar populations in the LUVIT galaxies to be estimated.    
In cases where there is no discernible stellar population associated with the target galaxy, the parallel imaging can be used to provide an estimate of the foreground and background populations present in the primary pointings.
In the \marthago\ program, parallels in F606W and F814W were obtained for the 6 galaxies (7 targets) where the ACS field fell within $3r_{25}$ (DDO210, IC1613, Sextans~A, Sextans~B, UGC6817, and both pointings of WLM).  
The location, orientation, and dither pattern of the parallel observations were set by the location, orientation, and dithers of the primary pointings.  

In \karriego, the F606W and F814W filters were each observed for one full orbit for every target, obtaining a full dither pattern in each filter.  For three-orbit targets, the third orbit was split between F606W and F814W, with an exposure in each filter obtained at each dither position. As with the primary observations, the exposure times were determined using the orbit planner in APT, maximizing the total exposure time within the orbital visibility window.
The resulting exposure times produce deep optical CMDs.  

A similar observational strategy was followed for the parallels obtained in \marthago.  ACS/WFC parallels were obtained in conjunction with the WFC3/IR observations.  For the targets with one WFC3/IR orbit (IC1613, Sextans~B), the parallels were split between the F606W and F814W filters, with longer exposures on F814W.  For the targets with at least two WFC3/IR orbits (DDO210, UGC6817, Sextans~A, both WLM positions), F606W and F814W were observed in parallel for one orbit each. UGC6817 had three WFC3/IR orbits; in this case, as in \karriego, the third orbit was split evenly between F606W and F814W.

\section{Simultaneous Multiband Data Reduction}\label{sec:data_redux}
  
This section summarizes the multiband LUVIT data reduction, including all \karriego, \marthago, and relevant archival imaging (Section~\ref{sec:archival_datasets}) for each target.  The LUVIT reductions use the photometric reduction pipeline developed to produce simultaneous multiband photometric measurements for the PHAT survey \citep{williams2014phat} and refined for the PHATTER survey \citep{williams2021phatter}. This pipeline uses the DOLPHOT package \citep{hstphot} to detect sources and perform point-spread-function (PSF) based fitting on those sources.  The following subsections provide a high-level overview of the reduction steps, highlighting any differences between the LUVIT reductions and the PHAT and PHATTER reductions.  Readers are referred to 
\citet{williams2021phatter} for further details of the reduction pipeline.  

Section~\ref{sec:data_redux_image_processing} describes the preprocessing steps which prepare the individual images for reduction and align the multiband images onto a common astronomical reference frame.  
Section~\ref{sec:data_redux_photometry} summarizes the procedure for obtaining simultaneous, $N$-band photometric measurements.
Section~\ref{sec:data_redux_catalogs} discusses the contents of the resulting stellar catalogs and the criteria used to identify stellar objects and assess the quality of individual photometric measurements. 

\subsection{Initial Image Processing and Alignment}\label{sec:data_redux_image_processing}
 
The data reduction pipeline starts from \texttt{flc} files for WFC3/UVIS and WFC/ACS observations, \texttt{flt} files for WFC3/NIR observations, and \texttt{c0m/c1m} files for WFPC2 observations, all downloaded from MAST. 
First, the images in each band from each individual visit are run through {\tt astrodrizzle} to update the data quality extensions to improve cosmic ray masking.  Then, each individual image is masked and calibrated for distortion using the DOLPHOT preparation tasks.  After these preparation steps, PSF photometry is measured separately on each individual detector from each individual exposure.  The resulting catalogs for all of these individual detector runs are then cross-matched across all of the exposures to refine the image world coordinate system of all of the headers \citep{williams2014phat}, aligning all of the images covering the region to a precision of $\sim$0.1 pixel.  Once the headers of the original \texttt{flc}, \texttt{flt}, and \texttt{c0m} images have been updated with the aligned world coordinate systems, the preparation process, including {\tt astrodrizzle} and the DOLPHOT preparation tasks, are rerun separately for each band but including all visits in the {\tt astrodrizzle} stack to optimize cosmic ray masking. 

For each target, a drizzled reference image was created from all exposures in the chosen reference camera and filter combination.  For the majority of the LUVIT reductions (all the targets included in \marthago), the drizzled image created from the exposures obtained in F475W (from either WFC3/UVIS or ACS/WFC, depending on the target) was specified to be used as the reference image.  The F475W band is the most likely to contain a large number of both red and blue stars, allowing it to be used to successfully align NIR images as well as NUV images.  For all \marthago\ targets, the F475W imaging was sufficiently deep to provide good alignment. 

Three targets in \karriego\ were not included in \marthago\ and used a different filter for the reference image. The camera and filter used for creation of the reference image for these targets were WFC3/UVIS F275W (DDO6, UGC8760) and ACS/WFC F555W (M81-DWARF-A). 
If not 
specified (as was done for the targets included in \marthago), the drizzled image produced from the filter with the greatest total exposure time is chosen by the pipeline as the default reference image.  For DDO6 and UGC8760, this was the drizzled F275W image. Without the need to align both NIR and NUV imaging, this resulted in successful multiband alignment for DDO6 and UGC8760. For M81-DWARF-A, the most successful alignment was found using the drizzled F555W image as the reference image, rather than the pipeline's default choice of F814W. 

Finally, we note that, as shown in Figure~\ref{fig:footprints}, the star-forming region of many of the LUVIT targets fits primarily within one WFC3/UVIS chip. In one case (DDO6), a lack of NUV sources on the one WFC3/UVIS chip that does not include the star-forming region resulted in a failure to find any acceptable astrometric solution between the UV and optical bands for that chip.  As a result, the reduction pipeline was run on only one of the two WFC3/UVIS and ACS/WFC chips (chip 2 in both cases).  Therefore, the photometric catalogs for DDO6 include only data from the WFC3/UVIS and ACS/WFC chips containing the star-forming region.  In general, the relative alignment between individual exposures and the drizzled reference image will have larger relative alignment uncertainties in regions of very low source density, where few stars may be matched between the individual image and the reference image.   

\subsection{Measuring Resolved Stellar Photometry}\label{sec:data_redux_photometry}

After the initial image processing and alignment described above, all images covering the region are put into DOLPHOT together to obtain, over the area covered by the bounds of the reference image (Section~\ref{sec:data_redux_image_processing}), full-depth photometry catalogs that include measurements of all stars in all images, as well as the resulting combined measurement for every measurement in each band.  
When DOLPHOT runs on a full list of aligned images with multiple bands including multiple cameras, it generates a stack of all of the image pixels in memory and searches the stack for peaks at least 3-$\sigma$ above the noise in the full stack including all bands.  It then forces a fit of the point spread function at the locations of all these peaks in every individual image, accounting for all neighbors.  It does this iteratively by fitting the stars, subtracting them from the images, and searching the residuals for stars missed in the previous pass.  It also compares the flux measured with PSF and aperture photometry for the most isolated stars in the data to correct the photometry for PSF-related systematics. The result is a measurement of the flux (or upper limit) of every source in the catalog from every image in the stack, as well as a combined measurement of the flux in each filter, taking advantage of multiple exposures in each band to refine the flux.  Table~5 lists the values of the user-specified DOLPHOT parameters utilized in the LUVIT reductions. 

\subsection{Photometric Catalogs}\label{sec:data_redux_catalogs}

The pipeline produces the standard, comprehensive, DOLPHOT output catalogs (`{\tt.phot}' files; described in detail in the DOLPHOT documentation\footnote{http://americano.dolphinsim.com/dolphot/}).  These catalogs include, for each source detected, the measured fluxes from the PSF fit for that source's location in each image, 
the calculated Vega system magnitude, 
the associated photon-count-based uncertainties, the signal-to-noise ratio, and metrics which indicate the quality of the PSF fit for that source (including \texttt{sharpness}, \texttt{roundness}, \texttt{crowding}, and $\chi$). Combined measurements for each source, in each filter, are also included in the DOLPHOT output catalogs.

We subsequently cull these photometric catalogs using the GST quality criteria established by \citet{williams2014phat}, and summarized in Table~\ref{table:gst_criteria}.  First, sources must be detected with a signal-to-noise ratio $>4$.  Second, we apply camera-specific quality criteria to the combined measurements in each filter, using the metrics related to the quality of the PSF fit.  The limiting values for these measurements were chosen to balance the goals of retaining a very high fraction of total measurements while removing a high fraction of the sources that lie outside the stellar features in the CMDs \citep{williams2014phat}.  

Specifically, we remove likely cosmic rays and background galaxies using the \texttt{sharpness} parameter, which provides a measure of the relative concentration of flux in the central vs.\ outer pixels compared to the PSF. High \texttt{sharpness} values (high central concentration) are typically caused by cosmic rays or hot pixels, while low values typically indicate background galaxies or blended stars.  
The \texttt{crowding} parameter provides a measurement of the extent to which a source's PSF fit was affected by neighboring sources. Larger \texttt{crowding} values indicate a higher density of other sources in the region of the measured source's PSF profile, and a higher likelihood that the photometric measurements have higher systematic uncertainties caused by imperfect subtraction of neighboring sources' PSF fits. Finally, $\chi$ provides a measurement of the goodness-of-fit of the PSF to the source's flux distribution.  

In addition to these GST quality criteria, we also remove detections that are associated with diffraction spikes, which are artifacts resulting from saturation of foreground bright stars. We do this by identifying areas with a local stellar surface density that exceeds the median stellar surface density of the immediately surrounding regions by $>$3--5$\sigma$, where $\sigma$ is defined as the expected Poisson noise (the square root of the local stellar surface density). The 3--5$\sigma$ threshold prevents us from removing bona-fide stars in crowded star-forming regions, while effectively eliminating objects associated with diffraction spike features.  The threshold was set on a target-by-target basis after visually inspecting the resulting sources identified for removal as diffraction spike artifacts.  We use a 3$\sigma$ threshold for NGC 4163, SagDIG, and WLM Position 1 and a 5 $\sigma$ threshold for Leo A, Sextans A, Sextans B, and WLM-POS2; we use a 4$\sigma$ threshold 
for the remainder of the targets. 

\begin{figure}
    \centering
    \includegraphics[width=\columnwidth]{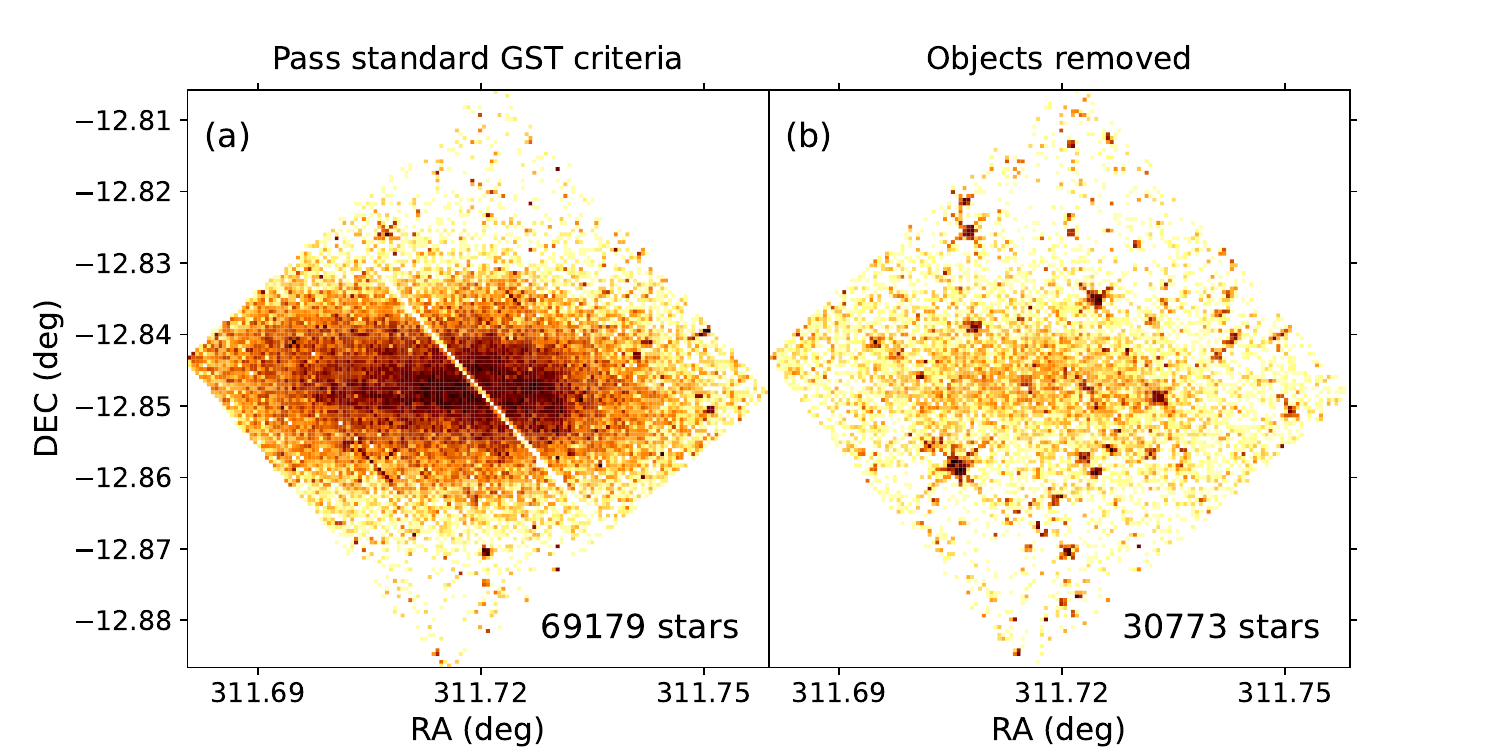}
    \includegraphics[width=\columnwidth]{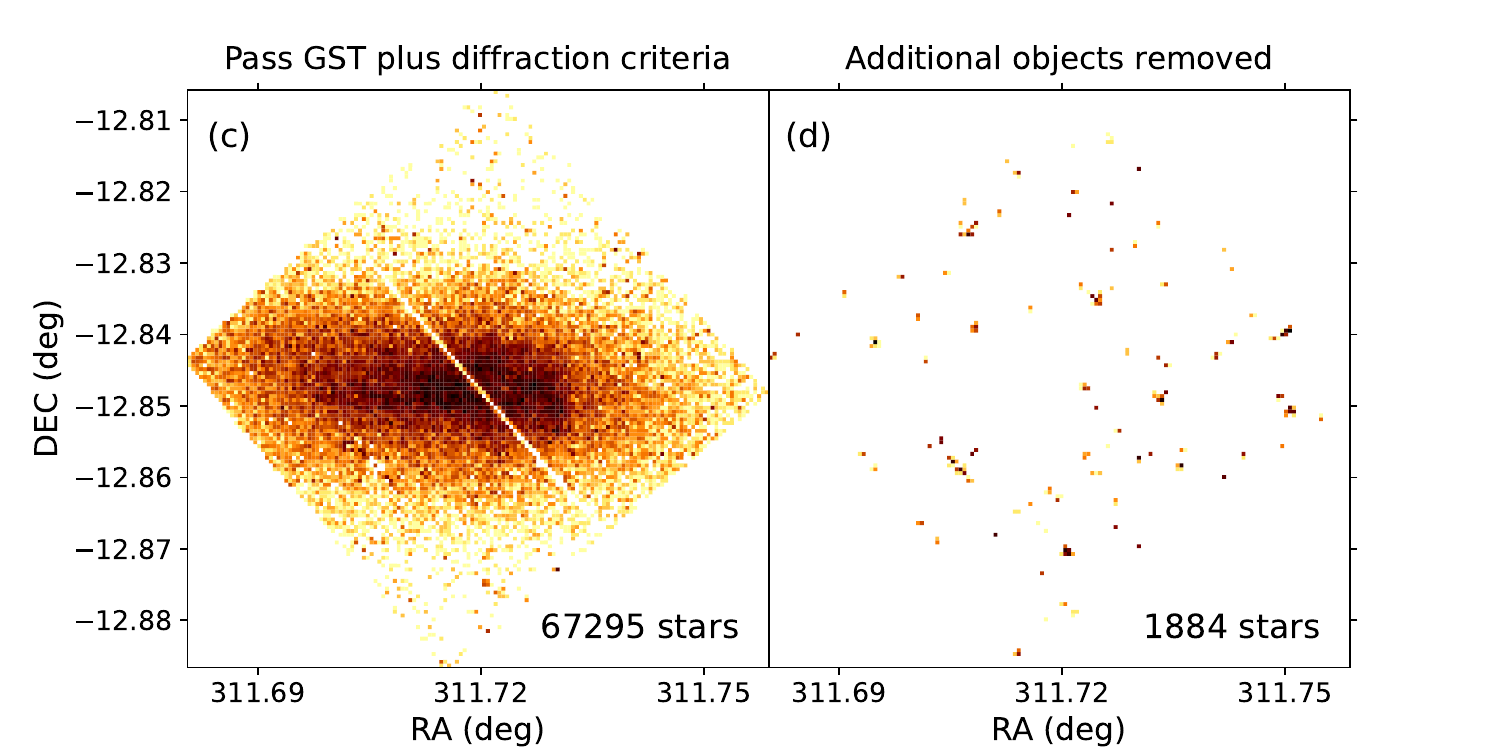}
    \includegraphics[width=\columnwidth]{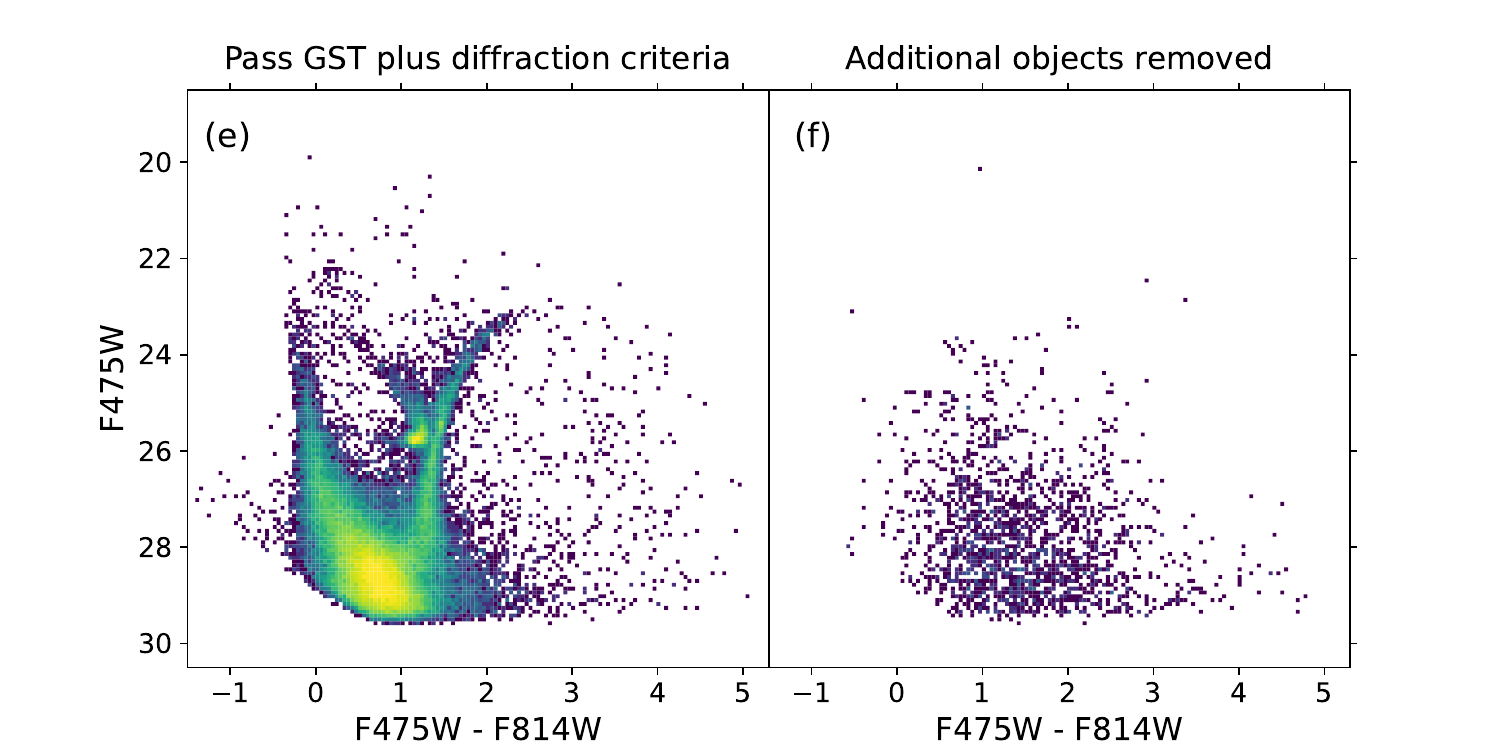}
    \caption{Spatial distribution (panels a-d) and color-magnitude distribution (panels e, f) of sources detected in the DDO210 archival optical observations.  The spatial distributions show sources which (a) pass the pipeline quality criteria for high quality stellar sources (GST); (b) do not pass the GST criteria and are therefore removed from the stellar sample; (c) pass both the GST criteria and an additional set of criteria designed to remove spurious detections associated with diffraction spikes; and (d) sources which are removed from the sample because they do not pass the diffraction spike removal criteria, but which do pass the GST criteria   (Section~\ref{sec:data_redux_catalogs}).
    The CMDs show sources that (e) pass both the GST and diffraction spike criteria and (f) are removed because they do not pass the diffraction spike removal criteria (same detections as in panel d).  While the standard pipeline quality (GST) criteria remove many of the spurious detections associated with diffraction spikes, the additional diffraction spike criteria we have implemented targets remaining spurious detections, while not removing sources that are likely associated with the stellar population of the target galaxy.}
    \label{fig:gst_v_dgst_catalogs}
\end{figure}

The above cuts are demonstrated in Figure~\ref{fig:gst_v_dgst_catalogs}, which shows the spatial position of sources in DDO210 that pass and fail the GST criteria, as well as the additional diffraction spike artifact criterion, and compares the optical CMDs of sources that pass all quality criteria to those that are removed as likely diffraction spike artifacts. 

Given the large range of wavelengths covered by the NUV to NIR observations, we do not require a source's measurements to pass the criteria simultaneously in all filters. Rather, we assess these criteria using the combined measurements for the filter(s) being analyzed.  Unless noted otherwise, all data presented in this paper pass both the GST quality and diffraction spike artifact criteria in the relevant filter(s).  

\section{Overview of LUVIT Stellar Catalogs}\label{sec:data}

\subsection{Spatial Distribution of Stellar Sources}\label{sec:spatial_dist_dgst}

 \begin{figure*}
    \centering
    \includegraphics[width=\textwidth]{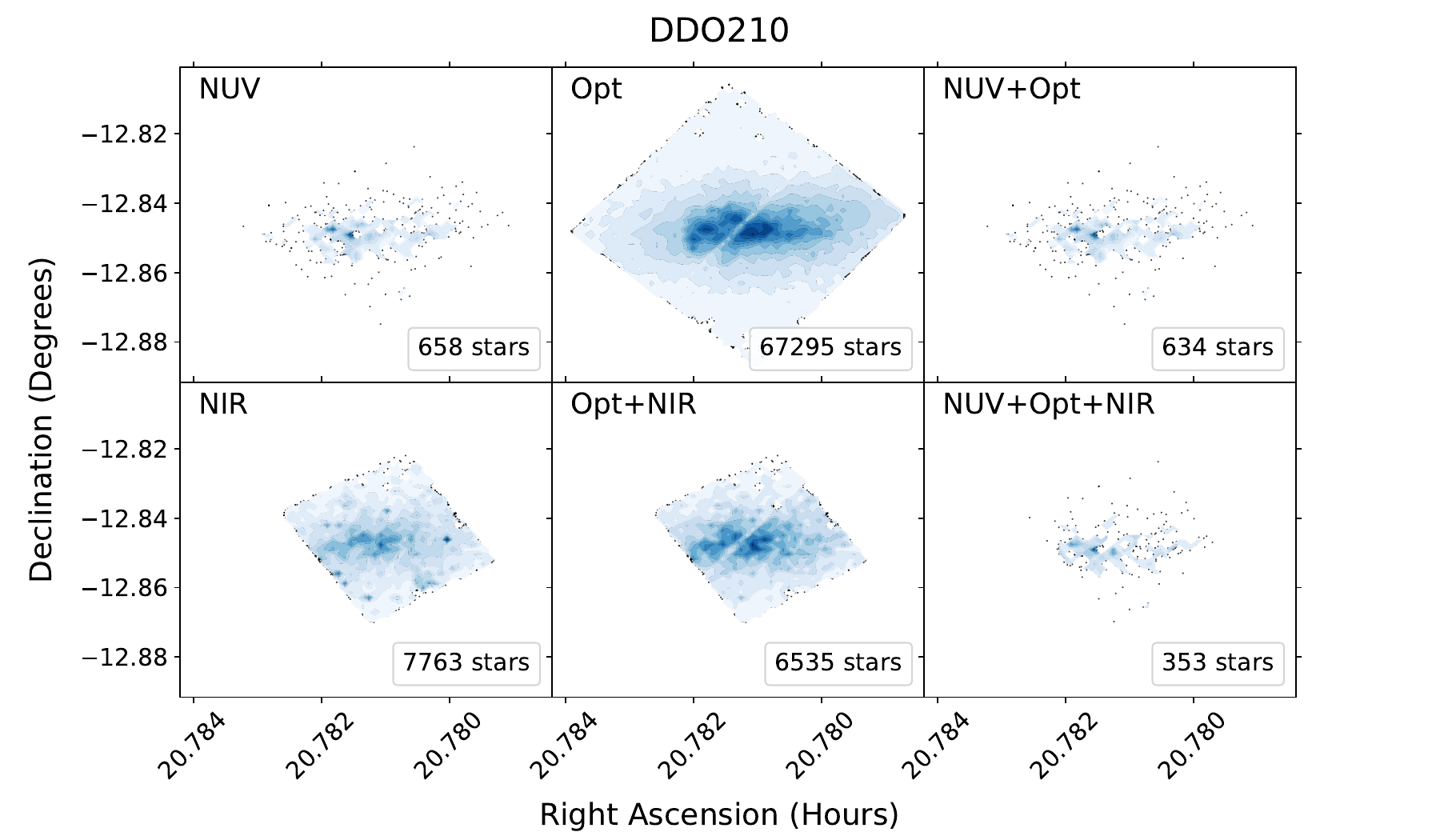}
    \caption{
    Distribution on the sky of stars passing all quality criteria (Section~\ref{sec:data_redux_catalogs}) in the NUV, optical, and/or NIR filters in DDO210, discussed in Section~\ref{sec:spatial_dist_dgst}. (Top) The spatial distribution of stars that meet quality criteria in two UV filters (left), at least two optical filters (middle), and in two UV and at least two optical filters (right).  (Bottom) The spatial distribution of stars that meet quality criteria in the two broadband LUVIT NIR filters (left), in the two NIR and at least two optical filters (middle), and in two UV, two broadband NIR, and at least two optical filters (right).  Individual stars are shown as points in the areas of lowest density.  
    Regions of higher stellar density are shown with small spatial bins color coded according to the number of stars, with darker blues indicating regions of higher spatial stellar density.  The color scale is determined independently in each panel in order to provide maximum contrast for seeing features in the spatial distribution of stars.  The complete figure set (23 images) is available in the online journal and includes a figure for each LUVIT pointing.  
    }
    \label{fig:stellar_spatial_dist}
\end{figure*}

Figure~\ref{fig:stellar_spatial_dist} 
shows the distribution on the sky of sources passing all stellar quality criteria (Section~\ref{sec:data_redux_catalogs}) in NUV, optical, or NIR (when available) filters, and in various combinations thereof.  
The maximum spatial extent of the stellar catalogs is set by the extent of the drizzled reference image used for each target (Section~\ref{sec:data_redux_image_processing}), 
and is largest when the drizzled reference image is based on ACS/WFC observations.  

The diversity in spatial scale of the LUVIT galaxies, relative to the HST instruments' fields of view, can be seen in the spatial distributions of the observed stars. More distant and/or compact galaxies, for which the LUVIT observations cover the entirety, or the vast majority, of the galaxy, show a clear drop in stellar density over the field of view of the LUVIT observations.  In other targets, typically those within the Local Group, the LUVIT pointings cover only a fraction of the galaxy and the stellar density remains relatively constant throughout.  

In the galaxies where the field of view of the HST observations is large enough to contain the bulk of the stellar population of the galaxy, the relatively compact spatial distribution of stars passing our quality criteria in both NUV filters reflects the fact that the majority of the most recent SF in each galaxy is 
found in the innermost regions.  Older RGB stars, detected in the optical and NIR, are more numerous and have a larger spatial extent than the brightest, young stars observed in the NUV.  The optical images are typically the deepest images, reaching further down the luminosity function than other observations, and providing greater sensitivity to low surface brightness populations.  
While old red giant branch stars are the dominant population of stars detected in the NIR bands, the NIR images are typically shallower than the optical images, with a smaller field of view (Figure~\ref{fig:footprints}).

\subsection{Color-magnitude Diagrams}\label{sec:data_cmds}

 \begin{figure*}
    \centering
    \includegraphics[width=\textwidth]{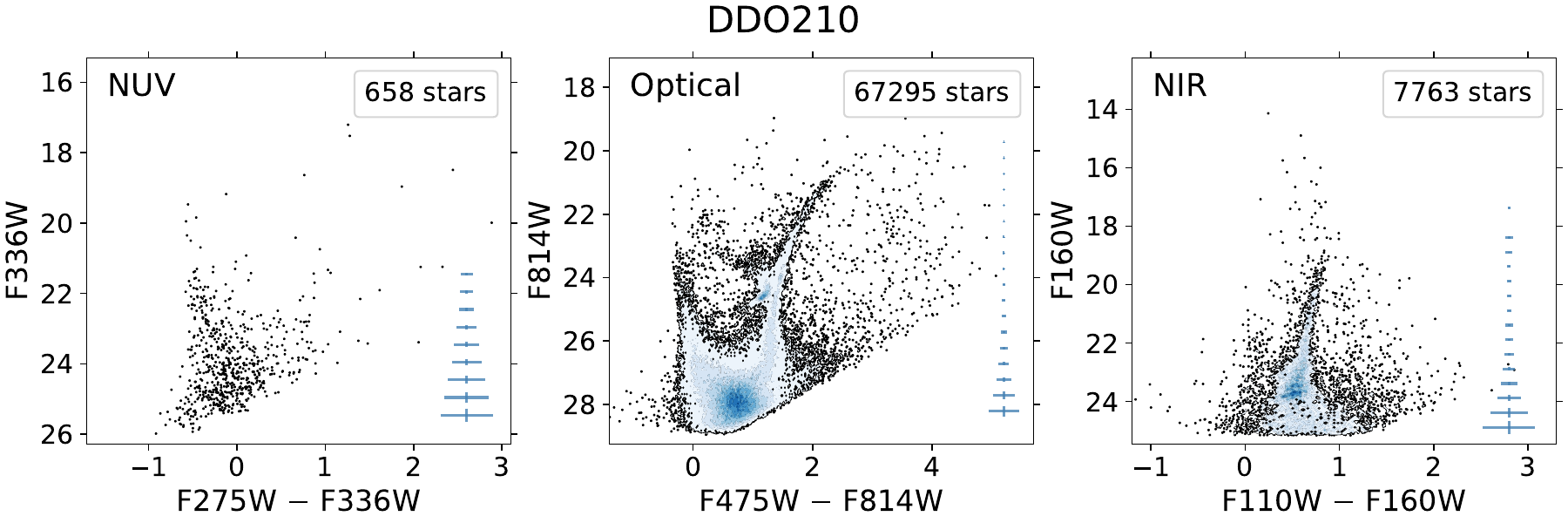}
    \caption{  
    CMDs of stars in DDO210 passing all quality criteria (Section~\ref{sec:data_redux_catalogs}) in the shown UV, optical, or NIR broadband filters, discussed in Section~\ref{sec:data_cmds}. Mean photometric uncertainties (based on photon counts) in color and magnitude are shown as a function of magnitude.  Regions of higher stellar density in the CMD are shown as a Hess diagram.  The color scale
    is determined independently in each panel in order to provide maximum contrast for seeing features in the CMDs, with darker blues indicating regions of higher density.  For targets with observations in more than two optical filters, multiple CMDs are shown, each using the reddest filter for the color baseline, demonstrating the increased separation between CMD features obtained with larger color baselines. 
    The magnitude ranges on the ordinate axes are set by the individual datasets, and are not the same across targets or filter combinations, in order to better show details in the CMDs.  The CMDs all show massive MS stars, and in many cases strong HeB sequences, both of which are indicative of significant recent SF, as expected for the star-forming LUVIT galaxy sample.
    The complete figure set (23 images) is available in the online journal and includes a figure for each LUVIT pointing.  
    }
    \label{fig:cmds}
\end{figure*}

 \begin{figure*}
    \centering
    \includegraphics[width=\textwidth]{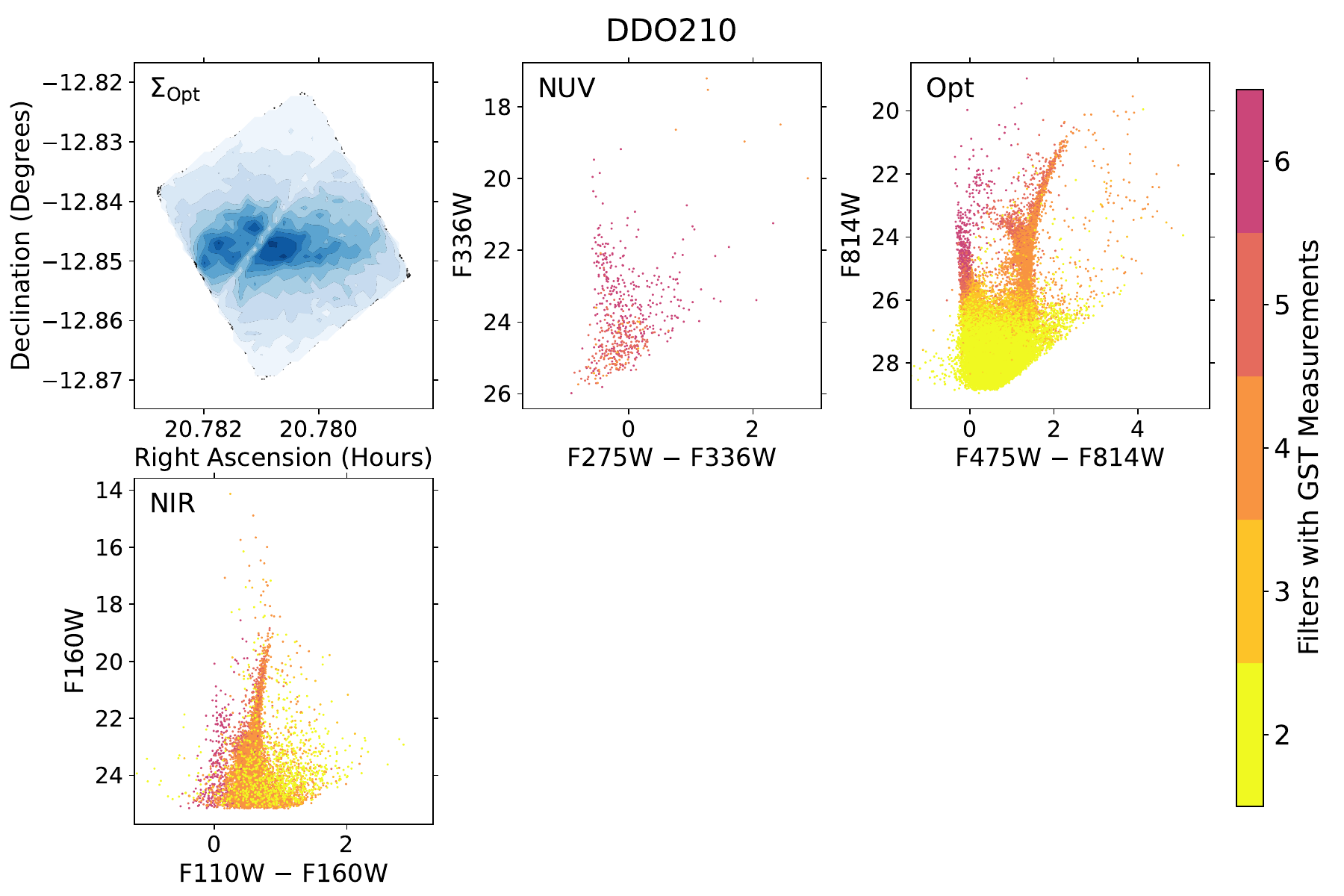}
    \caption{
    Location in various CMDs of stars in DDO210 passing all quality criteria (Section~\ref{sec:data_redux_catalogs}) in the shown UV, optical, or NIR broadband filters, in the area of sky covered by all filters (discussed in Section~\ref{sec:data_cmds}). (Leftmost panel:) 
    A representation of the area on the sky with full-filter coverage, showing the relative stellar density (as in Figure~\ref{fig:stellar_spatial_dist}) of stars that pass all quality criteria in at least two optical filters and are also within the spatial region where measurements in all filters are possible. 
    (Remaining panels): Each star is color coded by the total number of NUV through NIR broadband filters in which it has a measurement that passes all quality criteria.  As in Figure~\ref{fig:cmds}, for targets with more than two optical filters multiple optical CMDs are shown; however in this figure we instead show adjacent filter combinations for targets with more than two optical bands. Only stars that are located in a spatial area with full-filter coverage (leftmost panel) are shown in the CMDs.  The NUV detections are primarily the most massive (youngest) MS and BHeB stars in each galaxy, and are typically observed in the majority of the LUVIT broadband filters. 
    The complete figure set (23 images) is available in the online journal and includes a figure for each LUVIT pointing. 
    }
    \label{fig:cmds_nfilts}
\end{figure*}

\begin{figure*}
    \centering
    \includegraphics[width=\textwidth]{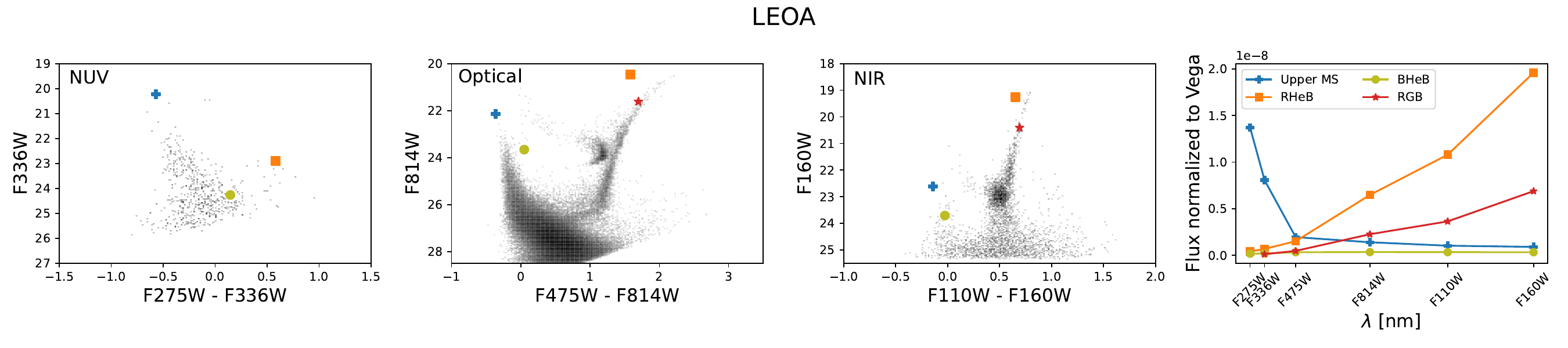}
    \includegraphics[width=\textwidth]{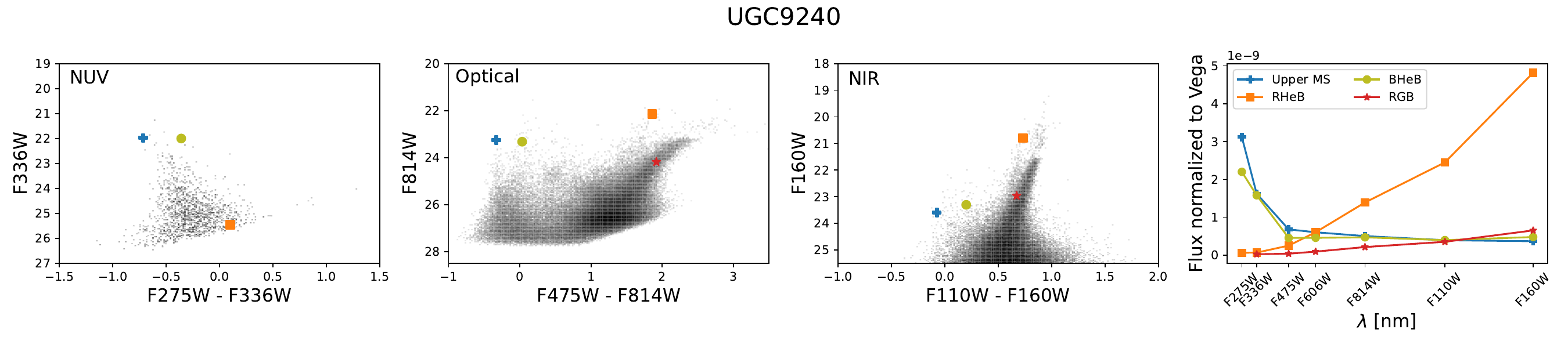}
    \caption{The location of stars in various stages of stellar evolution in NUV, Optical, and NIR CMDs (first three panels), along with their measured flux rates (when $>0$) in all available bands (right-most panels), for two LUVIT galaxies (Section~\ref{sec:data_cmds}).  Leo~A is amongst the lowest metallicity LUVIT galaxies, while UGC~9240 is amongst the highest.  Shown are the CMD locations and flux as a function of wavelength for a bright, massive MS star (blue crosses), a BHeB star (olive circles), a RHeB star (orange squares), and an RGB star (red stars; not detected in F275W).  The SEDs of the stars can be fit to stellar and dust models to estimate stellar parameters and dust properties along the line of sight to each star, as described in Section~\ref{sec:science_apps}.  
    }
    \label{fig:pedag_cmds_SEDs}
\end{figure*}

 \begin{figure*}
    \centering
    \includegraphics[width=\textwidth]{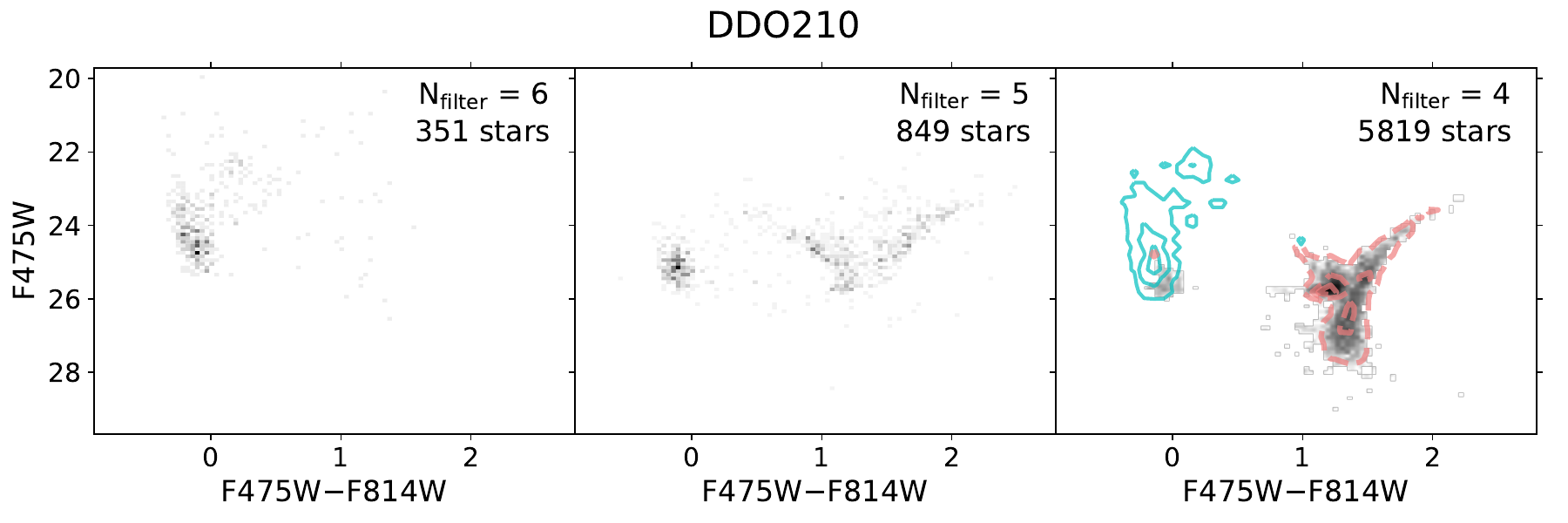}
    \caption{
    Optical binned CMDs of stars passing our quality criteria (Section~\ref{sec:data_redux_catalogs} in DDO210 in $N_{\rm filter}=6$ (left), $N_{\rm filter}=5$ (middle), and $N_{\rm filter}=4$ (right) broadband filters, as described in Section~\ref{sec:data_cmds}.  Stars included here must be detected in the two deepest optical filters, as well as two or more NUV or NIR filters.  Only stars located in the spatial area with full-filter coverage (as shown in Figure~\ref{fig:cmds_nfilts}) are included. 
    In the right-most panel, contours show the optical CMD distribution of stars detected in two UV and two optical (solid cyan contours) and in two optical and two NIR (dashed coral contours) filters.  The grayscale of the binned data is set independently in each panel, with a linear normalization in the left and middle panels and a logarithmic normalization in the right panel, in order to provide maximum contrast for seeing features in the CMDs. Darker values represent higher stellar counts.   
    In general, stars observed in all six filters are the most massive MS and BHeB stars, stars observed in five filters are lower-mass MS and BHeB stars as well as RHeB stars, and stars observed in four filters are typically MS and BHeB stars (UV and optical) and RGB, AGB, and RHeB stars (optical and NIR). 
    The complete figure set (20 images) is available in the online journal and includes a figure for each LUVIT pointing with \marthago\ or archival broadband NIR imaging.
    }
    \label{fig:cmds_yumi}
\end{figure*}

In this section, we present CMDs highlighting different aspects of the panchromatic LUVIT stellar catalogs.  Only sources meeting all quality criteria (Section~\ref{sec:data_redux_catalogs}) are shown in the CMDs.

Figure~\ref{fig:cmds} shows CMDs in the NUV, optical, and NIR for each target, derived from the full stellar catalogs.  For targets with more than two optical (or NUV) broadband filters, multiple optical (NUV) CMDs are shown.  To demonstrate the increased separation between stellar populations with increased color baseline, when multiple CMDs in the optical (or NUV) are shown, each uses the longest wavelength filter when computing the color and for the magnitude. 
A sequence of massive, young stars is seen in each of the NUV and optical CMDs.  Many LUVIT galaxies also display strong He-burning sequences (notably IC1613, Leo~A, NGC3741, NGC4163, UGC7577, UGC8508, UGC8651, UGC9240, and WLM), as well as significant AGB populations.  All display strong features associated with older stellar ages, such as the red giant branch, and for deep enough optical observations, the horizontal branch and red clump.  

The optical images typically reach the faintest apparent magnitudes, while the NUV and NIR images are shallower. Due to the heterogeneous nature of the archival data, a few targets with particularly complex archival datasets (Section~\ref{sec:notes_select_targets}) have limiting depths in the optical filters that vary significantly depending on the spatial position of a star.  For targets with particularly large differences in exposure times in a given filter as a function of sky position, this results in clear artifacts in the CMDs. The most egregious example is in Sextans~A, in the F555W and F814W (WFPC2) filters.

Most LUVIT targets are well removed from the MW disk plane (with absolute Galactic latitudes $b\gtrsim 40$\degree; Table~\ref{table:galaxysample}), and are thus expected to have relatively low MW foreground contamination in their CMDs.  The lowest Galactic latitude LUVIT galaxies are SagDIG ($b=-16.2879$), Antlia ($b=22.3124$), DDO~210 ($b=-31.341$), and M81~Dwarf~A ($b=33.0119$). In particular, the CMDs of SagDIG show significant foreground contamination; in addition to having a low Galactic latitude, SagDIG is also in the direction of the Galactic center.

Figure~\ref{fig:cmds_nfilts} presents, for each target, individual stars in each CMD color coded by the total number of filters in which the star passes all quality criteria.  To provide additional information content in this figure, for targets with more than two optical (or NUV) filters we plot one CMD for each of the adjacent optical (NUV) color combinations.  
Only stars in spatial regions of the sky with full-filter coverage (Section~\ref{sec:full_filter_region}; Figure~\ref{fig:cmds_nfilts}, left-most panel) are included in these CMDs.  For most targets, this is limited by the field of view of the WFC3/IR observations, and is a significant fraction of the footprints of the UV and optical observations (Figure~\ref{fig:footprints}).  However, for targets SEXTANS-A and WLM-POS-1, the heterogeneity of the archival datasets result in an unusual, and highly restricted, spatial area with full-filter coverage.  For WLM-POS-1 the effect is so severe that instead of showing the true full-filter region, a ``full wavelength coverage'' region is instead shown.  This region is computed by requiring observations to be possible in all filters with the exception of F439W, which is the filter which most severely restricts the spatial area (Section~\ref{sec:notes_select_targets}). 

Nearly all stars passing the quality criteria in both UV filters are also detected in at least one other broadband filter. In fact, most stars detected in both UV filters have measurements in the majority of the available filters. The stars with measurements in the greatest number of filters preferentially populate the upper MS and BHeB sequences. Brighter RGB stars and AGB stars are reliably observed in all optical and NIR filters.  A significant population of stars at the faint end of the CMDs are detected in only the optical or the NIR bands.

Figure~\ref{fig:pedag_cmds_SEDs} shows, for two LUVIT galaxies, the location in UV, optical, and NIR CMDs of individual stars in four stages of stellar evolution (upper end of the MS, BHeB, RHeB, and RGB), as well as the flux rates measured in each observed filter for each of the stars.  Note that due to the nature of the forced photometry (Section~\ref{sec:data_redux_photometry}), a flux rate is measured at the location of an observed source in all images (and can be negative). However, the rightmost panel of Figure~\ref{fig:pedag_cmds_SEDs} only displays a flux rate measurement 
when it translates into a detection passing all quality criteria.  Notably, for the RGB stars that do not appear in the NUV CMDs due to negative flux measurements in F275W, we do not plot their flux rate measurements in F275W in the right-most panels.

Figure~\ref{fig:cmds_yumi} demonstrates more clearly which stellar populations are typically observed in the optical and NUV, in the optical and NIR, or in all three wavelength regimes.  Figure~\ref{fig:cmds_yumi} shows the optical CMD distribution of stars, within the full-filter region, whose measurements pass all quality criteria in four or more filters ($N_{\rm filter}\ge 4$) for each of the targets with broadband NIR imaging.  To be included in Figure~\ref{fig:cmds_yumi}, the star must be measured in the two deepest optical filters for that target, with the additional measurements required to be in either the LUVIT NUV and/or the NIR filters. Thus, for a star shown in Figure~\ref{fig:cmds_yumi} as observed in four filters, the two nonoptical measurements may be in the NUV or in the NIR, or one may be in each.

The stars observed in at least six filters (required for Figure~\ref{fig:cmds_yumi} to be the two deepest optical, two NUV, and two NIR broadband filters) are typically the brightest and most massive upper MS and BHeB stars.  
The stars with measurements in five filters (required for Figure~\ref{fig:cmds_yumi} to be those that are observed in the two deepest optical and any three of the four LUVIT NUV and NIR broadband filters) are primarily the fainter (less massive) MS and BHeB stars, as well as RHeB stars. The majority of stars with measurements in four filters are observed in the optical and in either the NUV or the NIR (rather than in one of each of the NUV and NIR bands).  Contours in the right-most panel of Figure~\ref{fig:cmds_yumi} differentiate the CMD distributions of stars detected in two NUV and two optical bands (solid cyan contours; primarily MS and BHeb stars) and in two NIR and two optical bands (dashed coral contours; primarily RHeB, RGB, and AGB stars).

\section{Photometric Quality as Characterized by Artificial Star Tests}\label{sec:data_redux_asts}

We characterize the quality of the LUVIT photometry using artificial star tests (ASTs). We inject a series of ASTs of known properties at fixed sky positions into each overlapping input image and process the images through the photometric pipeline described in Section~\ref{sec:data_redux_photometry}, performing the same steps used for the observations. We compare the input and recovered properties of the artificial stars to quantify our photometric accuracy, precision, and completeness in each filter as a function of brightness and background stellar number density.  

The combination of multicamera, multiprogram data, as well as significant gradients in stellar density in many targets, results in a dependence of the photometric quality on spatial position for most targets. This makes quantitative analyses of many LUVIT targets complex. 
For the purposes of demonstrating the general quality of the photometry, the AST-based analyses presented below for each of the LUVIT targets are limited to the spatial area which contains observations in all filters with which that target was observed (hereafter referred to as the ``full-filter region'').  The one exception is WLM Position 1, for which the shallow observations in F439W were excluded from the determination of the full-filter region for this target, as discussed in Section~\ref{sec:notes_select_targets}.

\subsection{Determining the Full-filter Region}\label{sec:full_filter_region}
 
For the three targets without archival or \marthago\ NIR observations (DDO6, M81~Dwarf~A, and UGC8760), ASTs were run only in the full-filter region, which in these three cases provides good coverage of the optical extent of the galaxy. For these targets, the full-filter region was determined using a convex hull algorithm applied to the set of stellar measurements for which the photon count rate was nonzero in all filters.  The input AST lists were created only within the spatial boundary identified as the full-filter region.  

For the remainder of the LUVIT targets, ASTs were run over the full area of the reference image.  
In this case, the full-filter region was determined using a convex hull algorithm applied to the set of ASTs for which the count rate was non-zero in all filters.  For a subset of these targets (Leo A, Sextans A, Sextans B, UGC~6817, UGC~7577, UGC~8651, UGC~8833, WLM Position 1), the full-filter region was sufficiently complex that the convex hull algorithm was insufficient. In these cases, the spatial boundaries resulting from the convex hull algorithm were modified by hand to accurately capture features such as the corners of the WFPC2 ``sawtooth'' footprint (caused by the boundary between the Wide-Field Camera and Planetary Camera) and/or chip gaps that were not covered with dithered observations in one or more filters. Once the boundary of the full-filter region was established, all ASTs contained within the spatial boundary of the full-filter region were used for the AST-based analyses included below.

\subsection{Creation of Input AST Lists}\label{sec:input_asts}

For each target, input ASTs are generated simultaneously in all available bands using the capabilities provided by the Bayesian Extinction and Stellar Tool \citep[BEAST;][]{gordon2016}, which is an open-source project.\footnote{https://github.com/BEAST-Fitting} 
The BEAST maps a stellar evolution library onto a stellar atmospheric model based on an input stellar initial mass, age, metallicity, effective temperature, and surface gravity. To generate the list of input ASTs, we populate a physics model grid by varying stellar and dust parameters. We use PARSEC version 1.2S isochrones \citep{Bressan2012,tang2014,chen2015} for the stellar evolutionary library. For the stellar atmospheric models, we adopt the grid by \citet{Castelli2003} for local thermodynamic equilibrium (LTE) models (lower temperature stars; spectral type A and cooler) and the TLUSTY OSTAR and BSTAR grids by \citet{Lanz2003,Lanz2007} for non-LTE models (higher temperature stars). The BEAST provides implementation of a smoothly varying range of possible dust properties (from SMC-like to MW-like dust curves) via the amount of dust extinction ($A_{V}$), the average dust grain size ($R_{V}$), and the strength of the 2175\AA\, bump. Combinations of requested dust properties 
are applied to the intrinsic stellar spectral energy distributions (SEDs) computed from the stellar evolutionary library and stellar atmospheric models to compute extinguished SEDs.

For each galaxy, we first create a stellar physics model grid, with dust effects applied, that appropriately covers the galaxy's entire potential stellar and dust parameter space. 
We divide the range of fluxes in the physics model grid into 40--70 linear bins in magnitude space, depending on each galaxy's physics model coverage, leading to a typical flux bin size of 0.3\,--\,0.6 magnitude in each filter. 

To capture nonlinear changes in photometry characteristics with background source density, 
we also create stellar number density maps with 5\arcsec$\times$5\arcsec\ resolution. We use a catalog of all sources that pass the GST criteria in two select optical filters,\footnote{For the primary LUVIT reductions including \karriego, \marthago, and archival data, these filters are F475W and F814W.  Of the three targets not included in \marthago, the optical filters were F475W and F814W for DDO6, F555W and F814W for M81~Dwarf~A, and F606W and F814W for UGC8760.}
have DOLPHOT flags of 0 or 2 in those filters, and which fall within the region of full-filter coverage for that target. 
The stellar number density is computed in the reference image filter of each target over a magnitude range, typically between roughly 20$^m$--27$^m$, where the completeness is expected to be close to 100\%.\footnote{Since this magnitude range has to be determined before performing actual ASTs, we rely on our team's previous experience with similar HST data as well as the results of previous relevant studies \citep[e.g.,][]{dalcanton2009, choi2020}.  We confirmed the chosen range for each target was reasonable using the results of the full suite of ASTs.} 
We utilize 5--10 linear bins in stellar number density. The number of bins is dependent on the dynamic range of each galaxy, with 5 (10) stellar density bins used for a galaxy with the smallest (largest) dynamic range in stellar density. 

We randomly select model SEDs from the dust-applied stellar physics grid until at least 70 models are selected per flux bin and per stellar number density bin, in each filter. 
This first suite of input ASTs comprises roughly half of the final input ASTs of each galaxy, and is designed to cover the entire physics model grid (including areas well removed from the observed CMD space), which is required for the BEAST stellar SED fitting. 

To ensure adequate ASTs to support CMD modeling, we supplement the above with a higher density of ASTs within the region of observed CMD space. This supplement includes at least 4,000 (and up to 20,000) model SEDs per stellar density bin, focusing on the observed color and magnitude ranges for a given target. This ensures sufficient ASTs to compute completeness as a function of color and magnitude, which is essential for the CMD modeling. 

\begin{figure}
    \centering
    \includegraphics[width=\columnwidth]{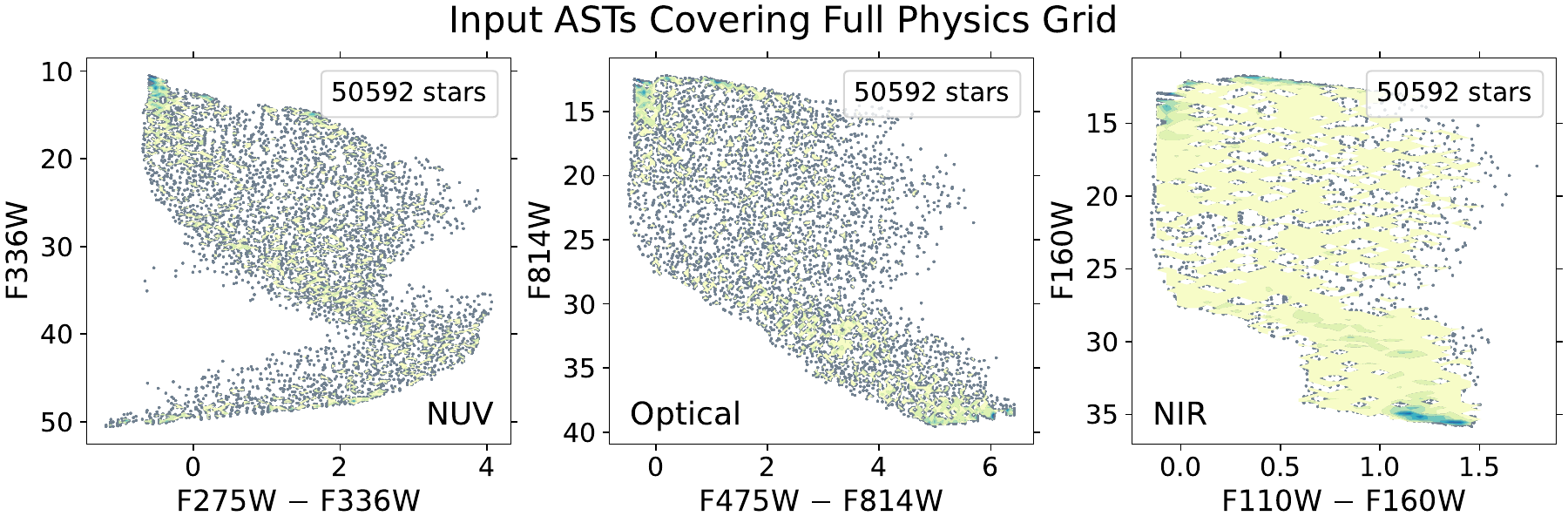}
    \includegraphics[width=\columnwidth]{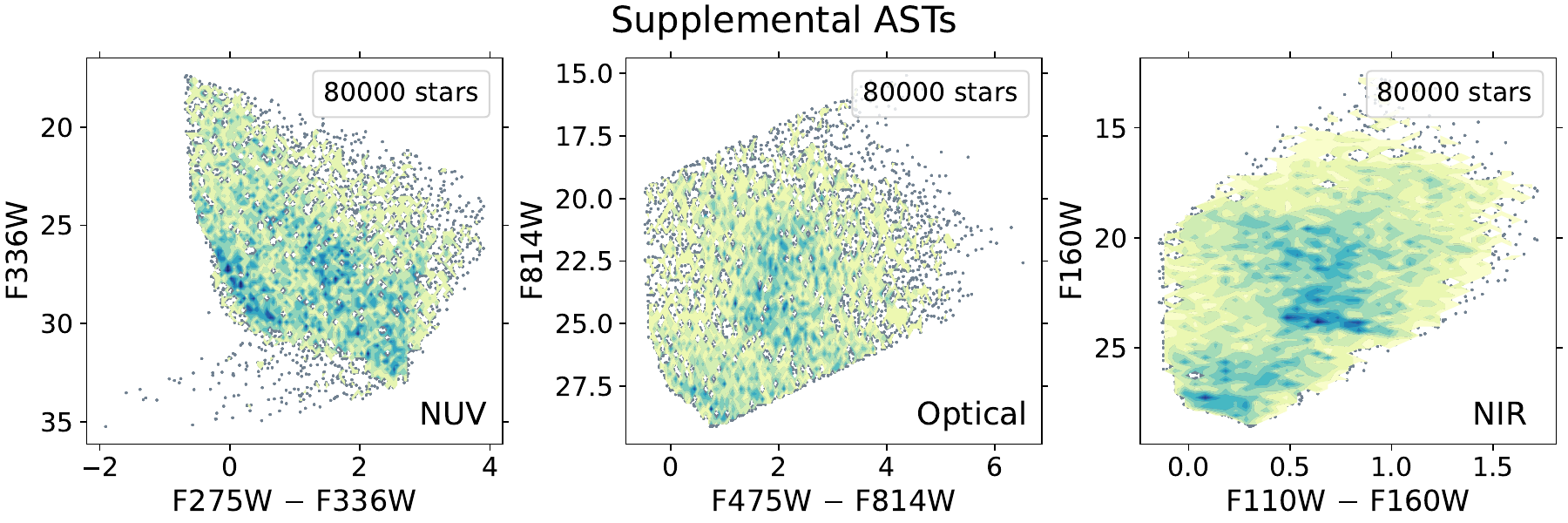}
    \caption{Location in CMD space of the input ASTs for 
    DDO210 (Section~\ref{sec:input_asts}).  (Top) Input ASTs are derived from a random selection of model SEDs from the full parameter space of the physics input grid, with extinction applied.  (Bottom) Supplemental input ASTs are derived from randomly selected model SEDs with a more limited apparent magnitude range and a higher percentage drawn from the region of observed CMD space.  ASTs are shown as points in regions of low density, and with a color scale increasing from yellow to green to blue (light to dark) in regions of higher density, with the range determined independently in each panel to provide maximum contrast.  
    }
    \label{fig:inputASTs}
\end{figure}

Figure~\ref{fig:inputASTs} shows the CMD location of the DDO210 input ASTs which cover the extinguished physics grid (top) and the supplemental ASTs which include a higher density of stars in the region of the observed CMD (bottom).  The number of flux bins, number of model SEDs per flux bin, number of stellar number density bins, and number of supplementing model SEDs per spatial density bin vary with the properties of the individual galaxies (e.g., dynamic range of stellar number density) and observations (e.g., depth). These numbers are empirically determined based on extensive tests and previous studies \citep[e.g.,][]{gordon2016, choi2020, VanDePutte2020, Yanchulova2021}.  
In general, this results in an input AST list that is a factor of $\sim$1.2\,--\,2$\times$ greater than the number of sources observed with a signal-to-noise ratio $>4$ in at least one filter. 

\subsection{Injection and Measurement of the ASTs}\label{sec:measure_asts}

The input AST list for each target, generated as described above, consists of assigned locations and magnitudes for each of the filters included in the target's dataset. Each of the entries in the input AST list is used to inject an artificial star into the observed images, and the same photometric measurement routine is run as described in Section~\ref{sec:data_redux_photometry}.  The artificial stars are injected one at a time, so that the measurement of a given artificial star is not affected by the others in the input AST list.  This procedure results in the same photometric measurements and quality metrics that are reported for the observed astronomical sources (Section~\ref{sec:data_redux_catalogs}).

For the following analysis, an artificial star is deemed to be ``recovered'' in a given filter if the absolute difference between the input and measured magnitudes is less than \astmagdiffthreshold\ magnitudes, the star is recovered within two reference pixels of its input location, and the photometric measurement of the star passes all GST quality criteria as described in Section~\ref{sec:data_redux_catalogs}.  ASTs placed in areas affected by diffraction spikes are not included in the computation of the recovery fraction, bias, or photometric uncertainty as a function of magnitude.

\subsection{Photometric Quality as a Function of Magnitude}\label{sec:data_results_asts}

For each target, we compare the DOLPHOT measurements of the injected ASTs to the input magnitudes to evaluate the recovery fraction of ASTs (Figure~\ref{fig:completeness}, Table~\ref{table:completeness}), as well as the accuracy (Figure~\ref{fig:bias}) and uncertainty (Figure~\ref{fig:phot_unc}) of the measured magnitudes, as a function of magnitude in each filter.  As discussed in Section~\ref{sec:full_filter_region}, we limit this analysis to the spatial region with full-filter coverage, with the exception of WLM Position 1 (Section~\ref{sec:notes_select_targets}).  

Significant spatial variations in the photometric quality are to be expected within a given filter.  These variations may be due to intrinsic changes in the stellar density within a galaxy, leading to varying levels of crowding impacts on the stellar measurements.  However, given the heterogenous nature of the archival observations, as well as the presence of chip gaps in the ACS, WFC3/UVIS, and WFPC2 imaging, spatial variations in the photometric quality can also be due to any combination of the following: observations obtained in the same filter but with different exposure times; partially overlapping observations in the same filter resulting in more total exposure time; or areas falling within a chip gap in a filter, resulting in less total exposure time.  Thus, the AST-based analyses that follow should be considered to be representative of the general quality of the LUVIT data for a given target.   

\subsubsection{Completeness}\label{sec:completeness}

 \begin{figure*}
    \centering
    \includegraphics[width=\columnwidth]{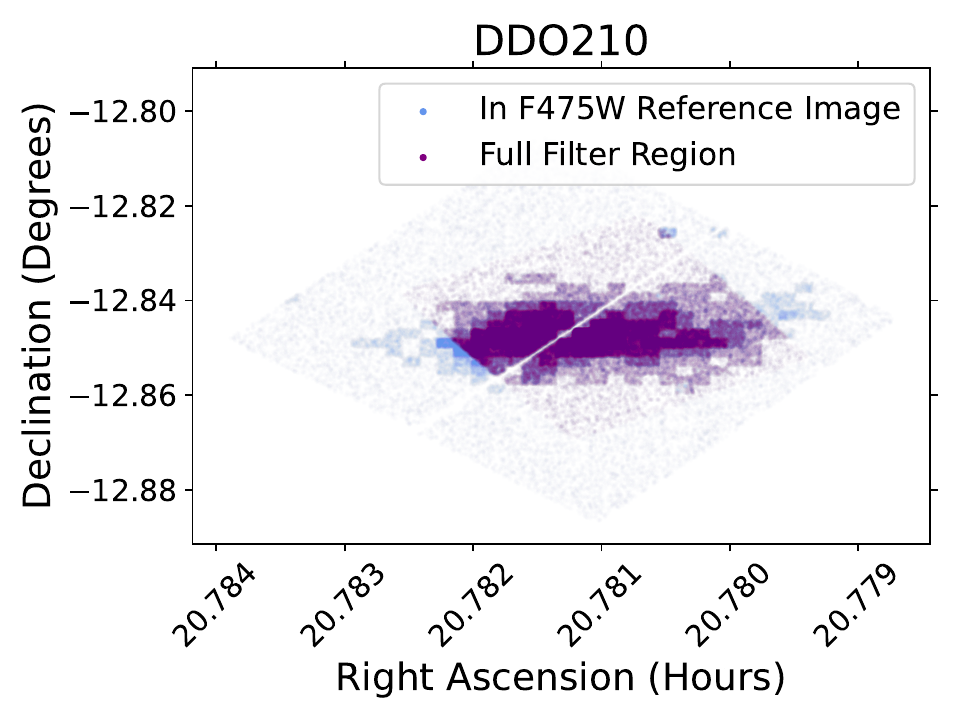}
    \includegraphics[width=\columnwidth]{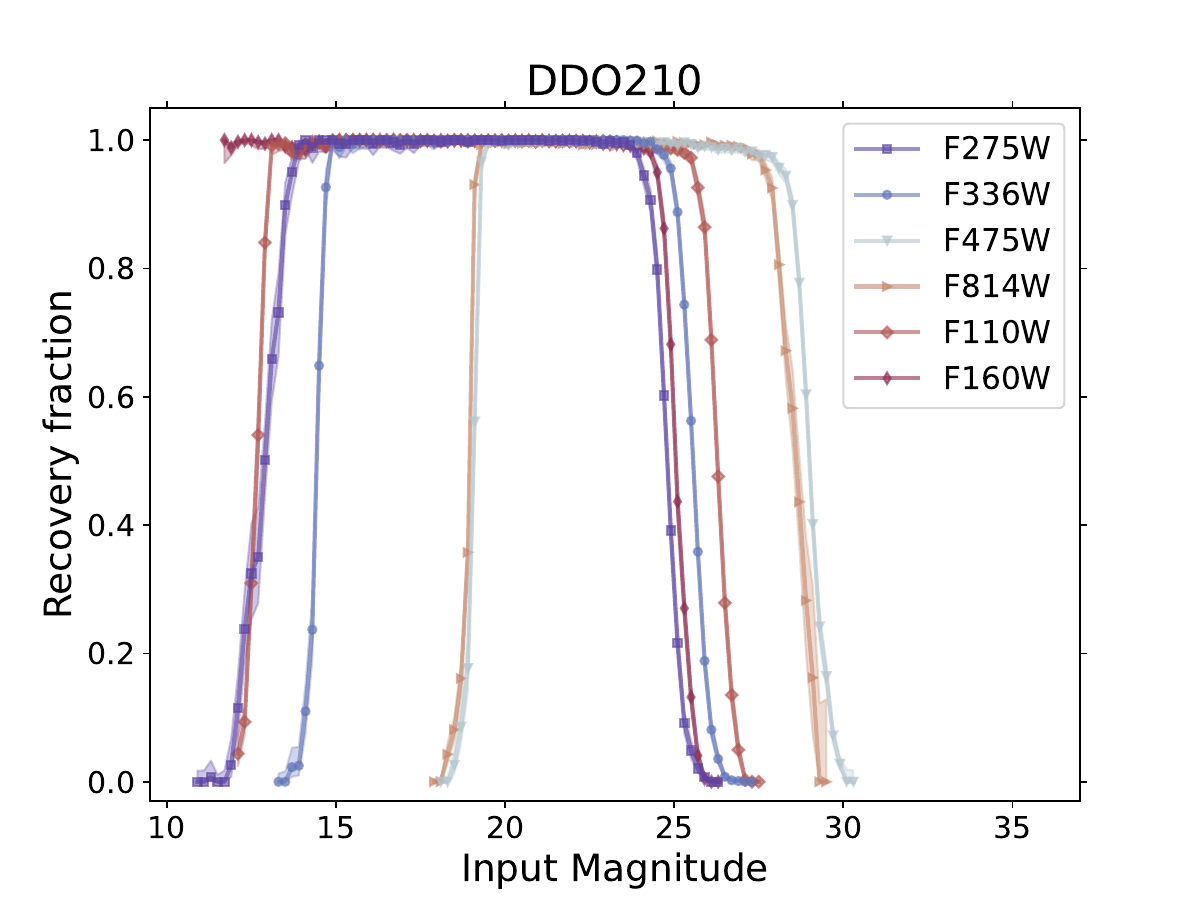}
    
    \caption{(Left) Location of artificial stars inserted throughout the DDO210 reference image (blue) and falling within the ``full-filter region'' (purple; Section~\ref{sec:full_filter_region}) used for calculating the recovery fraction and photometric uncertainty and bias presented in this and following figures (Figures~\ref{fig:bias} to \ref{fig:completeness_bias_spatial}). 
    (Right) Fraction of artificial stars recovered as a function of magnitude in each filter for DDO210 (Section~\ref{sec:completeness}).  The recovery fraction is computed in nonoverlapping bins of \CLmagbin\ magnitude.  An artificial star is deemed to be `recovered' if the absolute difference between the input and measured magnitude is less than \astmagdiffthreshold\ magnitudes, the star is recovered within two
    reference pixels of its input location, and the photometric measurement of the star passes all quality criteria as described in Section~\ref{sec:data_redux_catalogs}.  Saturation impacts the recovery fraction of stars at the brightest magnitudes. Shaded regions show the 95\% confidence interval on the recovered fraction of ASTs, computed using the Jeffrey's interval.  
    The complete figure set (23 images) is available in the online journal and includes a figure for each LUVIT pointing.  
    }
    \label{fig:completeness}
\end{figure*}

The fraction of ASTs recovered is computed in non-overlapping bins of \CLmagbin\ magnitude, with the 95\% confidence intervals on the recovered fraction of ASTs computed using the Jeffrey's interval\footnote{Calculated using the \href{https://docs.astropy.org/en/stable/api/astropy.stats.binned_binom_proportion.html}{binned\_binom\_proportion function} in the astropy stats package.}. Figure~\ref{fig:completeness} displays the results for all filters, as well as showing the location of the full-filter region relative to the reference image, and the spatial density of ASTs included throughout the reference image.  The confidence intervals shown in Figure~\ref{fig:completeness} only capture the statistical uncertainty in the recovery fraction as a function of magnitude. They do not include potential variation in the recovery fraction as a function of sky position within a filter, as discussed above.

Table~\ref{table:completeness} provides the magnitudes in each broadband filter, for each target, at which 50\%, 75\% and 90\% 
of the ASTs are recovered\footnote{Completeness levels for the \marthago\ medium-band observations will be discussed in Boyer et al. (in prep.)}.  The computed values in the \CLmagbin\ magnitude bins have been linearly interpolated to determine the completeness limits shown in Table~\ref{table:completeness}.  We caution that these should be treated as representative values over the full-filter area, and note that 
completeness limits in a given filter may differ significantly with spatial position, for the reasons discussed above.  This is explored further in Section~\ref{sec:spatial_dependence}. 

While the filters with the deepest observations are optical filters, the optical filters also have the greatest variation in depth.  Given the heterogeneous nature of the archival observations used in the LUVIT reductions, the magnitudes at which the recovery fraction deviates from 1 can vary significantly from target to target in the archival filters.  The depth of the \karriego\ NUV and the \marthago\ NIR and F475W observations are more uniform from galaxy to galaxy than the archival optical and NIR observations, due to the LUVIT observing strategy of obtaining a set number of orbits per target, depending on the distance of the target (and using only 2 distance bins in both the NUV and NIR; Section~\ref{sec:obs_strategy}). 
The F336W (F110W) observations are deeper than the F275W (F160W) observations by roughly $\sim 0.8$ ($\sim 1$\,--\,1.3)~magnitudes. The optical filters typically have the faintest saturation limits, with saturation often impacting the measurements at magnitudes $\gtrsim 17$. In contrast, saturation often only starts impacting the NUV and NIR measurements at magnitudes $\lesssim 15$. 

\subsubsection{Bias}\label{sec:bias}

 \begin{figure*}
    \centering
    \includegraphics[width=\textwidth]{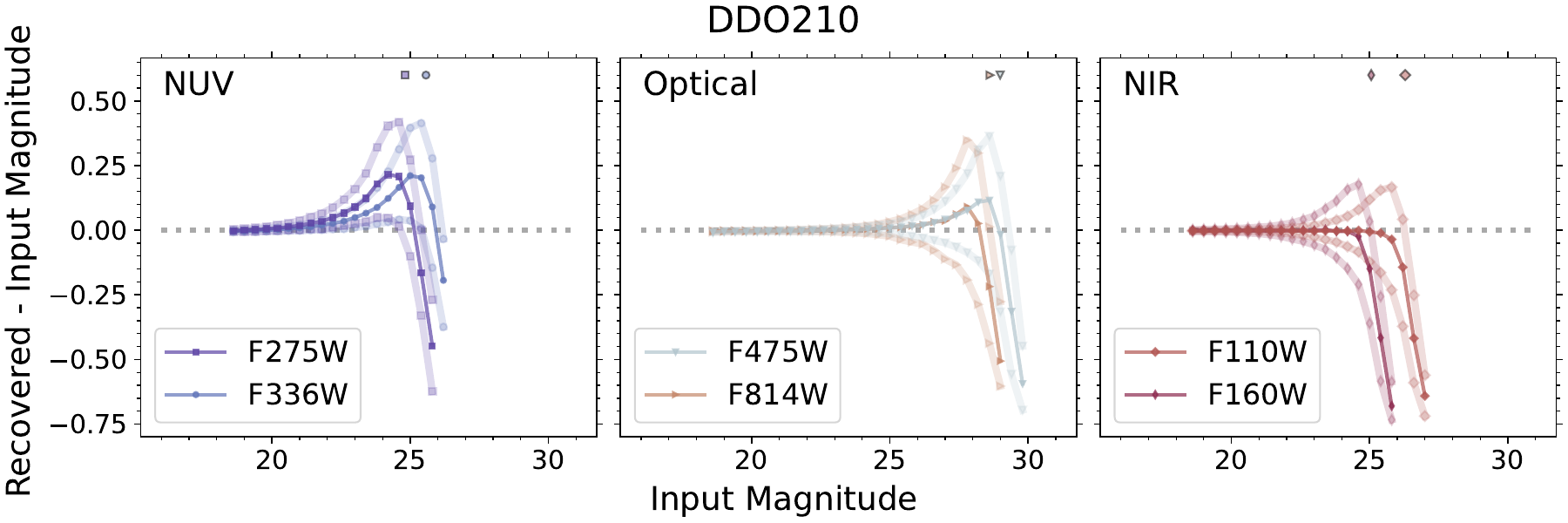}
    \includegraphics[width=\textwidth]{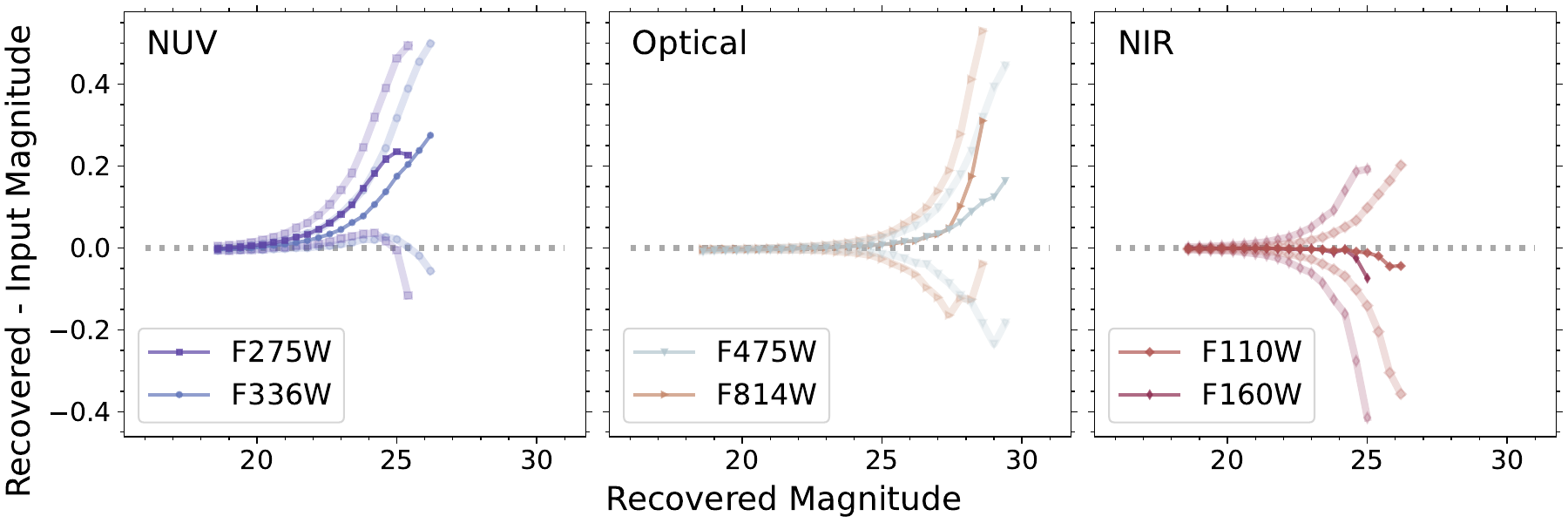}
    \caption{Photometric bias as a function of magnitude in each broadband filter for DDO210 (Section~\ref{sec:bias}), as a function of the input (top) and recovered (bottom) magnitudes of the ASTs. 
    The bias is computed in bins of \biasmagbin\ magnitude, for bins in which at least \minstarsbiasmagbin\ artificial stars are recovered.  
    The 50th percentile of the photometric bias distribution in each filter is shown with thin lines, and the 16th and 84th percentiles of the distribution are shown with the thicker, more transparent lines.  The 50\% completeness limits in each filter, which are computed as a function of input AST magnitude, are denoted by the corresponding symbols placed along the top edge of the top panels. The photometric bias is calculated in the spatial area with full-filter coverage. 
    The complete figure set (23 images) is available in the online journal and includes a figure for each LUVIT pointing.
    }
    \label{fig:bias}
\end{figure*}

To assess the bias in the measured magnitude of a star, the difference between the input and measured magnitudes in a filter is computed for every recovered AST.  The distribution of the difference in the measured vs.\ the input magnitude of ASTs is then assessed in bins of \biasmagbin\ magnitude, for all magnitude bins with a minimum of \minstarsbiasmagbin\ recovered ASTs.  Figure~\ref{fig:bias} displays, for each filter, the 16th, 50th, and 84th percentiles of the photometric bias distribution as a function of both input and recovered magnitude within the full-filter region.  
The representative 50\% completeness limits in each filter, in the full-filter region (Section~\ref{sec:completeness}), are also shown for reference as a function of input magnitude.

The NIR measurements are generally consistent with having no median bias while the recovery fraction remains close to 1.  At bright input magnitudes (generally $\lesssim 20$ in the UV, and variable in the optical), all NUV and optical filters also show no significant bias in the measurements.  A comparison of the median bias for fainter input magnitudes in the NUV and optical, down to the $\sim 50$\% completeness limit in each filter, show that the NUV measurements typically have the largest median bias, with recovered magnitudes fainter than the input magnitudes.  The median bias in the optical measurements are more heterogeneous; a few targets have little median bias while the recovery fraction remains close to 1, while others have a median bias intermediate between that of the NUV and NIR, with recovered magnitudes also generally fainter than the input magnitudes.  In all bands (NUV, optical, and NIR), the median bias trends steeply downward with increasing input magnitude
 (moving into the regime where the recovered magnitudes are brighter than the input magnitudes) when the recovery fraction starts to drop significantly below 1.  The overall widths of the bias distributions, as estimated by the difference between the 16th and 84th percentiles of the bias distribution in each filter, are generally roughly similar between the different bands. This is easiest to compare in the panels which plot the bias as a function of recovered magnitude in Figure~\ref{fig:bias}.  

The patterns in bias seen in the LUVIT observations are generally consistent with those seen in other resolved stellar observations of nearby galaxies, including the lower density regions of the PHAT, PHATTER, and PHAST surveys of M31 and M33 \citep[Chen et al., accepted]{williams2014phat, williams2021phatter}, and the Scylla survey of the LMC and SMC \citep{murray2024}.  At fainter apparent magnitudes as the luminosity function of stars is increasing, and in crowded regions of higher stellar density, stars that cannot be detected individually can blend with their neighbors.  The blended sources can rise above the detection threshold, which biases the measurement to brighter magnitudes, as seen in the NIR, some optical observations, and when the recovery fraction is low.  

The tendency to recover sources biased to fainter magnitudes, as seen in the UV and some optical observations, was attributed in the PHAT dataset to charge transfer efficiency (CTE) effects causing over-estimation of the sky brightness \citep{williams2014phat}.  However, this bias pattern continues to be seen in datasets, including LUVIT, PHATTER, PHAST, and Scylla, that use preflashed and CTE-corrected WFC3/UVIS images, which should mitigate the impacts of CTE on the estimated sky flux.  Regardless of origin, a bias toward fainter recovered magnitudes likely indicates that the sky flux in those bands is being overestimated.   

\subsubsection{Photometric Uncertainties}\label{sec:phot_unc}

 \begin{figure}
    \centering
    \includegraphics[width=\columnwidth]{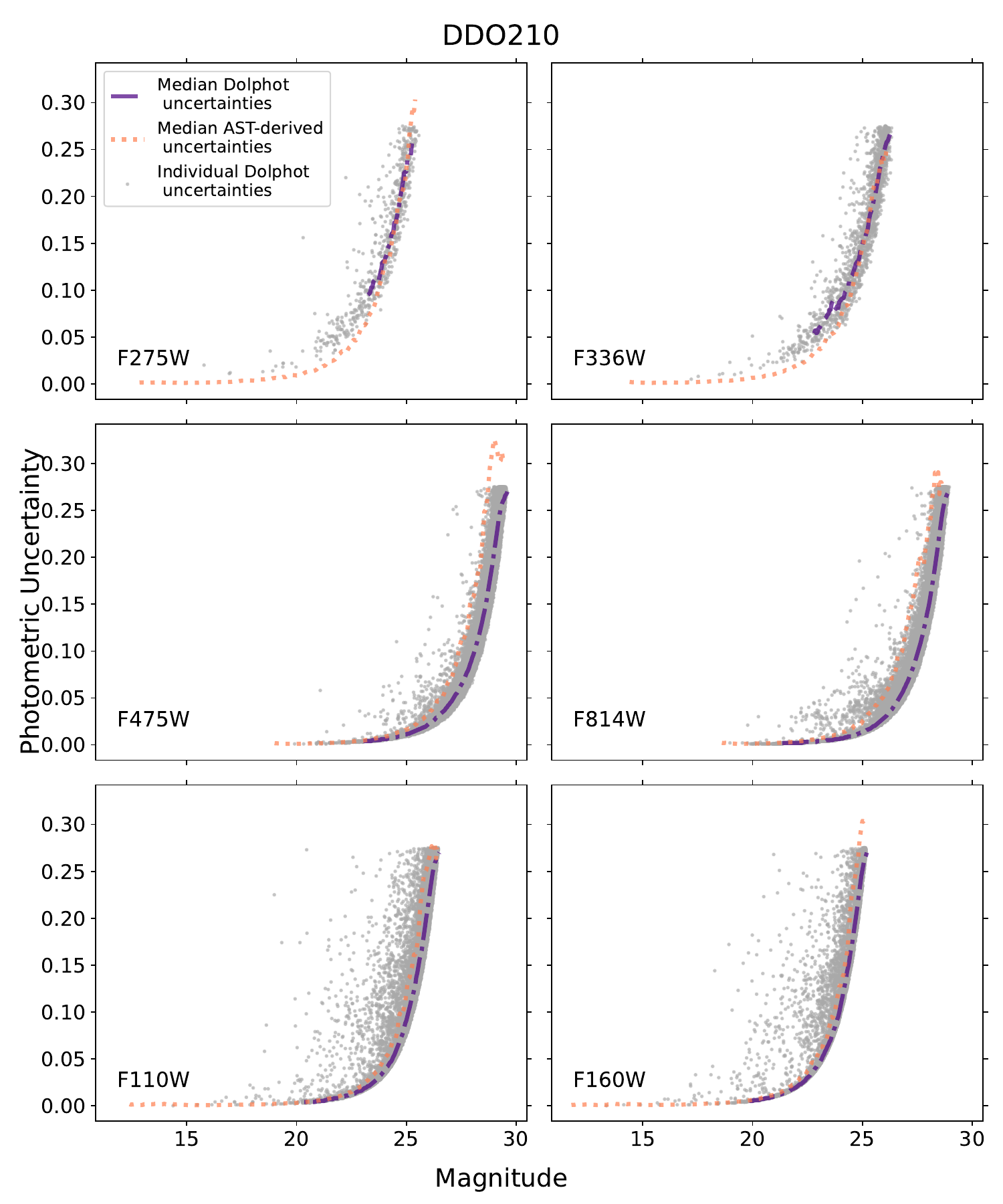}
    \caption{Photometric uncertainty as a function of magnitude in each broadband filter for DDO210 (Section~\ref{sec:phot_unc}).  Gray points show the individual DOLPHOT (Poisson-based) uncertainties for all stars passing the `dgst' criteria in a given filter (Section~\ref{sec:data_redux_catalogs}); these uncertainties are based only on photon-counting uncertainties.  The purple curve denotes the median Dolphot uncertainty as a function of measured magnitude, computed in running bins of \magbinphotunc\ magnitude, with a difference of \magbinphotuncbincenter\ magnitude between bin centers, and requiring a minimum of \minstarsphotuncmagbin~stars per bin.  The orange line shows the estimated uncertainty as derived from the ASTs as a function of recovered magnitude, defined as the standard deviation of the difference between the recovered and input AST magnitudes. The AST-derived uncertainty is computed using the same magnitude bins, but requiring \minastsphotuncmagbin~recovered ASTs per bin. The median DOLPHOT uncertainty and the AST-derived uncertainties are both computed using 3$\sigma$ clipping of outliers.  The ASTs capture sources of uncertainty beyond those due to Poisson statistics. One of the most significant is increased measurement uncertainty due to overlapping PSFs (`blending' of stars).  
    The complete figure set (23 images) is available in the online journal and includes a figure for each LUVIT pointing. 
    }
    \label{fig:phot_unc}
\end{figure}

The stellar measurements made by DOLPHOT include estimates of the uncertainty of the measured magnitudes.  These uncertainty estimates are based on the estimated S/N of the photometric measurement, which assumes Poisson statistics for the uncertainty in the source photon counts (as measured with the PSF-fitting algorithm) and the sky photon counts, and also includes the readout noise \citep{hstphot}.  The photometric uncertainties estimated by DOLPHOT in each filter as a function of magnitude are shown in Figure~\ref{fig:phot_unc}.  The median photometric uncertainty as a function of magnitude is also shown.  The photon-based uncertainties follow the expected trend of increasing with increasing magnitude.  

However, the photon-based DOLPHOT uncertainties do not capture all the potential contributions to the uncertainty in the measurement of a source.  One of the most significant additional sources of uncertainty in measurements of stars in nearby galaxies comes from the overlap of the PSFs of neighboring sources.  This occurs when the density of sources becomes sufficiently high that sources can no longer be treated as isolated sources, and is commonly referred to as stellar crowding.  While DOLPHOT employs algorithms specifically designed to account for overlapping PSFs of adjacent sources (Section~\ref{sec:data_redux_photometry}), the resulting measurements will in general have larger inherent uncertainties than is reflected in the photon-based uncertainties reported by DOLPHOT.

The ASTs can be used to estimate the true uncertainty in the stellar measurements as a function of magnitude, as they inherently account for the impacts of blended sources on the recovered magnitude (Section~\ref{sec:measure_asts}). Figure~\ref{fig:phot_unc} compares, as a function of recovered (measured) magnitude, the photometric uncertainty as derived from the ASTs to the uncertainties in the stellar measurements as reported by DOLPHOT.  The AST-based uncertainty is defined as the standard deviation of the difference between the input and recovered magnitudes of the ASTs. 

The inherent stellar surface density of a target is just one of the factors that determines the impact of blended sources on the measured magnitudes.  The level to which measurements are affected by crowding is also impacted by both the depth of the observations (deeper observations will generally result in a higher density of detected sources, amplified by the luminosity function of stellar sources increasing with increasing absolute magnitude) as well as the size of the PSF and pixel scale in a given image (wavelengths resulting in a smaller PSF will result in better separation between adjacent point sources, while smaller pixel scales typically enable higher fidelity modeling of the PSF).  

The LUVIT targets vary significantly in both their inherent maximum stellar densities and their distances, resulting in a significant range of observed stellar surface density.  LUVIT also covers a broad wavelength range, includes images from multiple instruments and channels, and has significantly different depths in different filters, especially in the archival optical imaging. Thus, the impacts of blended sources on the LUVIT measurements is expected to vary significantly and nontrivially by filter and by target.  This is evident in Figure~\ref{fig:phot_unc}, where the difference between the uncertainties as estimated by the ASTs and the median DOLPHOT photon-based uncertainties varies significantly between filters in a given target, and amongst targets.  
When the blending of sources has a significant impact on the photometric uncertainties, the AST-derived uncertainties will be systematically larger than the Dolphot-based measurement uncertainties.  This can be seen in the F475W and F814W filters in DDO210; these observations are deep, with $50\%$ completeness limits of $>28^m$ (Figure~\ref{fig:phot_unc}). 
In general, the impacts of stellar crowding on the photometric uncertainties appear largest in the optical filters (typically the deepest, and thus most crowded, observations), and smallest in the UV filters (typically the shallowest, least crowded observations).  The impacts of crowding on the photometric uncertainties are also apparent in many of the LUVIT targets' NIR broadband observations, due to the larger pixel size of the WFC3/IR detector.

\subsection{Spatial Dependence of Completeness and Bias}\label{sec:spatial_dependence}
As previously cautioned, all of the AST-based results reported above should be viewed as representative values, computed over the spatial region with full-filter coverage for each target.  Significant spatial variations in completeness, photometric uncertainties, and bias as a function of magnitude in the LUVIT targets can be caused by either the heterogeneity of the archival datasets and/or a significant range of observed stellar densities encountered across the field of view of the LUVIT observations, as discussed in Section~\ref{sec:data_results_asts}.  
Moreover, although the LUVIT targets are low-mass dwarf galaxies, 
the blending of sources does appear to impact the measurements in some filters in many of the LUVIT targets, as demonstrated in Figure~\ref{fig:phot_unc}, although the effect is minor for most targets, especially compared to the impacts seen in more massive galaxies \citep[e.g.,][]{williams2014phat}. 
In this section, we explore the impact of observed stellar surface density on the completeness limits and measurement bias.  

We use 5\arcsec\ by 5\arcsec\ spatial surface density maps, computed analogously to those computed for creating the input AST lists (Section~\ref{sec:input_asts}), but restricted to sources passing all stellar quality criteria (Section~\ref{sec:data_redux_catalogs}) and within the full-filter area.  For each target, we compute the 25th, 50th, and 75th percentiles of the distribution of stellar spatial densities in the reference image filter, creating four bins of stellar density for each target.  The upper limit of the lowest density bin is defined by the source density $\Sigma_{25}$ at which 25\% of the observed sources fall in a 5\arcsec\ by 5\arcsec\ source density map ``pixel'' with source density $\Sigma\le\Sigma_{25}$.   The upper limit of the second-lowest density bin is then defined by the source density $\Sigma_{50}$ at which 50\% of the observed sources fall in a region with source density $\Sigma\le\Sigma_{50}$, with $\Sigma_{25}$ forming the lower bound on the second-lowest density bin. The third and fourth density bins are defined in the same manner.
Thus, the limits of each bin are tied to the specific dynamic range and stellar density distribution of the observations (within the full-filter area). Each AST is then assigned to a stellar density bin based on the observed stellar density $\Sigma$ (again using the 5\arcsec\ by 5\arcsec\ spatial density map) at the spatial position of the injected AST.  We then compute, for each stellar density bin and for each filter, the fraction of recovered ASTs as well as the photometric bias.  This method ties the analysis of all filters to the stellar density distribution of the stellar sources detected in the reference image filter.

\begin{figure*}
    \centering
    \includegraphics[width=1.0\textwidth]{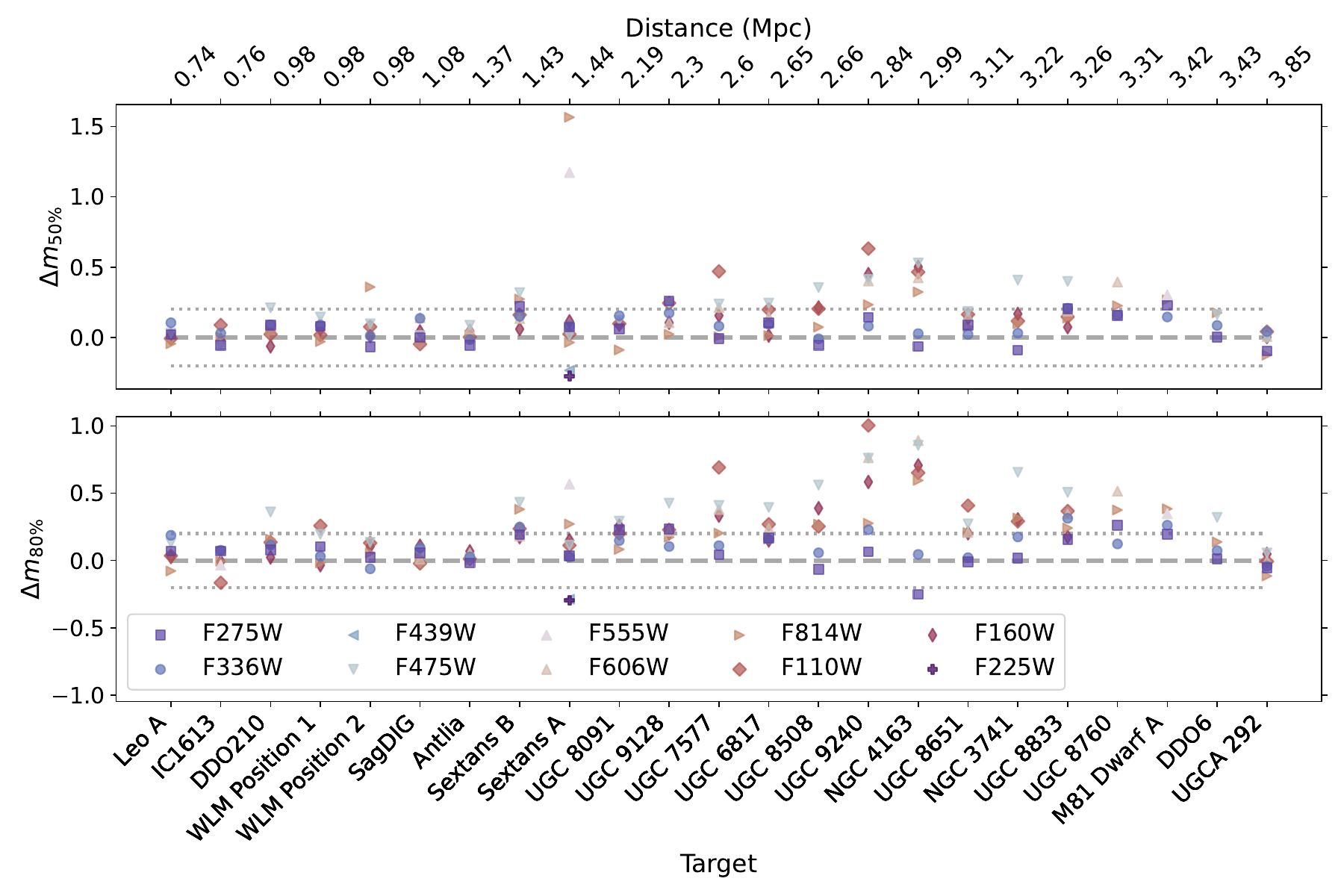}\caption{The difference in the 50\% (\textit{top panel}) and 80\% (\textit{bottom panel}) completeness limits calculated in the lowest ($\Sigma \leq \Sigma{_{25\%}}$) and highest ($\Sigma > \Sigma{_{75\%}}$) spatial density bins for each LUVIT pointing, for each filter (Section~\ref{sec:spatial_dependence}).  The pointings are presented in order of increasing distance of the target galaxy.  The lowest density bin is defined for each individual target by the source density $\Sigma_{25\%}$ at which 25\% of observed sources fall in a region with source density $\Sigma \leq \Sigma{_{25\%}}$.  The highest density bin threshold is defined as the source density $\Sigma_{75\%}$ at which 75\% of observed sources fall in a region with source density $\Sigma \leq \Sigma{_{75\%}}$. The dotted lines in each panel indicate a $\pm 0.2$~magnitude difference between the completeness limits in the lowest and highest spatial density bins.  The recovery fraction of ASTs and the bias in the recovered magnitudes as a function of recovered magnitude in four spatial density bins are shown in Figure~\ref{fig:completeness_bias_spatial} for targets which exceed the $|\Delta m|>0.2$ threshold in any filter, for either the 50\% or 80\% completeness limits. 
    }
    \label{fig:crowding_summary}
\end{figure*}

\begin{figure*}
    \centering
    \includegraphics[width=0.5\textwidth]{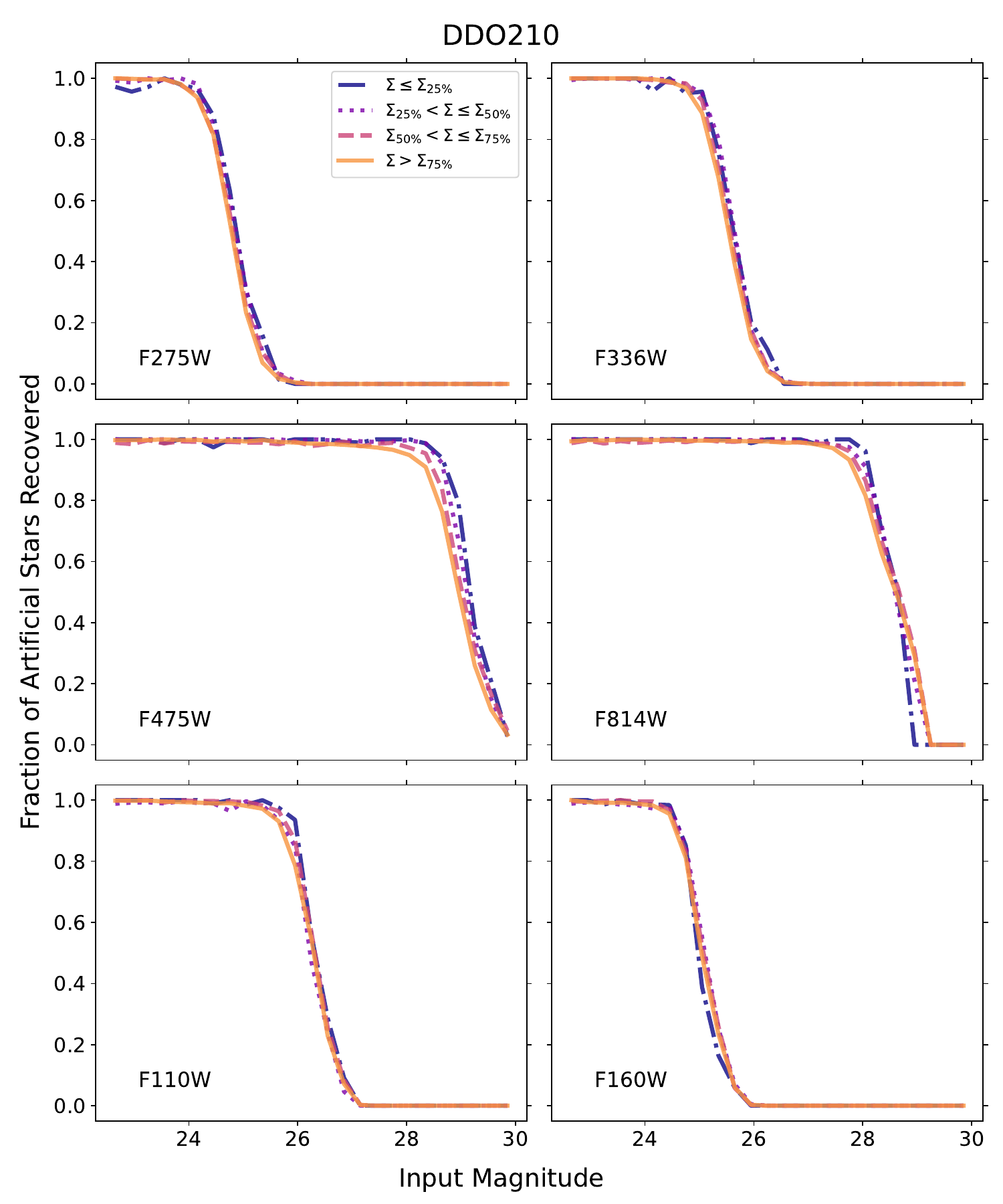}\includegraphics[width=0.5\textwidth]{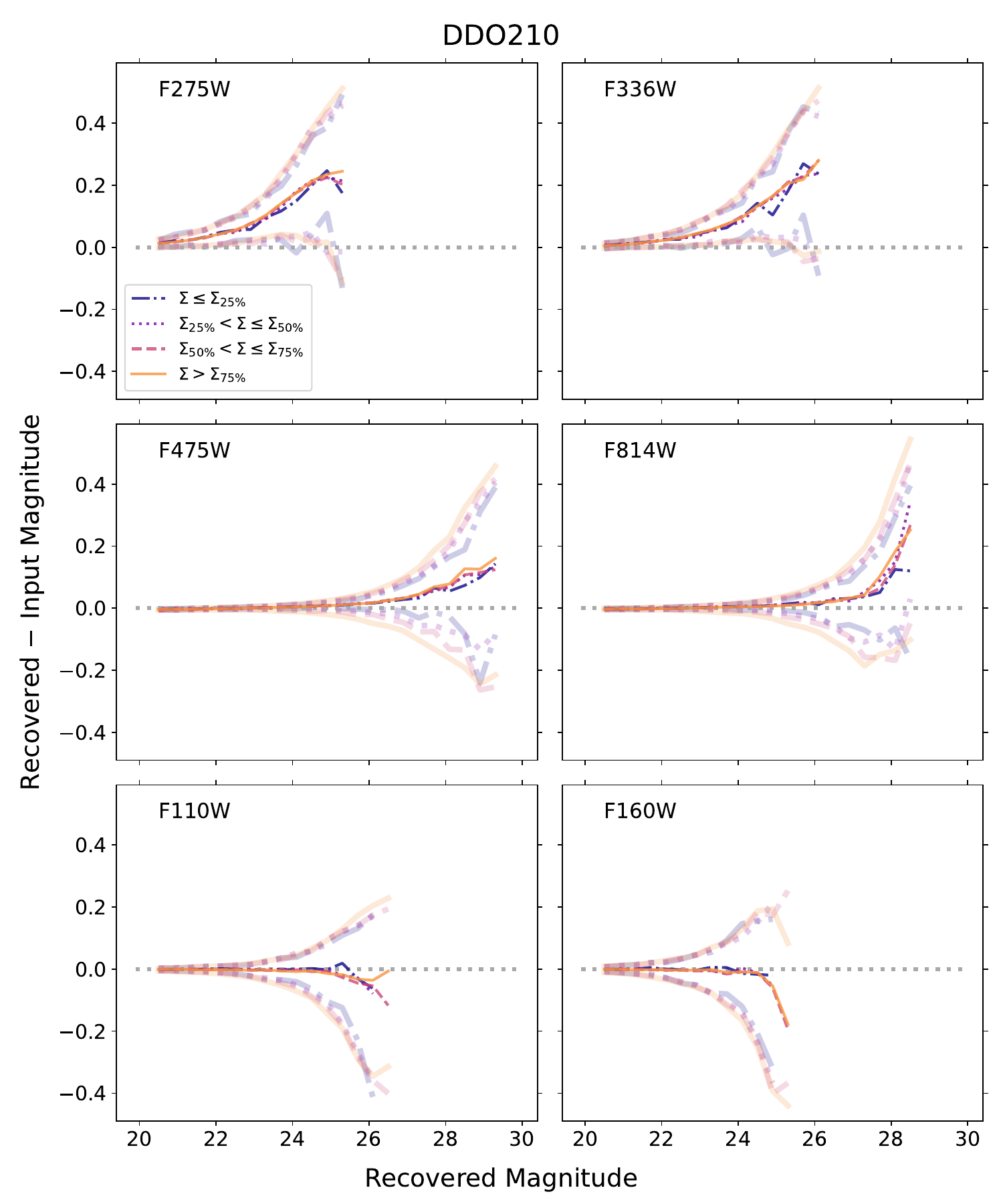}
    \caption{Recovery fraction and bias as a function of spatial
        density (Section~\ref{sec:spatial_dependence}). (\textit{Left
        panels}) Fraction of artificial stars recovered as a function
        of magnitude in each filter for DDO210, in bins of spatial
        density.  The spatial density bins are defined, per target,
        by percentiles (e.g., the lowest density bin is defined by
        the source density $\Sigma_{25\%}$ at which 25\% of observed
        sources fall in a region with source density $\Sigma \leq
        \Sigma{_{25\%}}$), as described in
        Section~\ref{sec:spatial_dependence}. The recovery fraction
        is computed in non-overlapping bins of width \spdclmagbin\
        magnitude. (\textit{Right panels}) Photometric bias as a
        function of recovered magnitude in each filter for DDO210,
        calculated in the same four spatial density bins shown in the
        recovery fraction (left) panels.  The bias is computed in
        non-overlapping bins of width \spdbiasmagbin~magnitude. The
        complete figure set (18 images) is available in the online
        journal, and includes a figure for each LUVIT pointing for
        which either $|\Delta m_{80\%}|$ or $|\Delta m_{50\%}|$ is
        $>0.2$~magnitudes in at least one filter in
        Figure~\ref{fig:crowding_summary}.
        \label{fig:completeness_bias_spatial}}
\end{figure*}

The results are summarized for all targets in Figure~\ref{fig:crowding_summary}, which shows the difference in the 50\% (top panel, $\Delta m_{50\%}$) and 80\% (bottom panel, $\Delta m_{80\%}$) completeness limits between the lowest and highest spatial density bins for each filter in each target.  We choose a threshold of $\pm 0.2$~magnitudes in $\Delta m$ to identify potential crowding impacts. Figure~\ref{fig:completeness_bias_spatial} displays the recovery fraction and bias as a function of magnitude (input and recovered, respectively) in each of the four spatial density bins for all targets that are identified as being potentially affected by crowding in at least one filter in Figure~\ref{fig:crowding_summary} (18 of the 23 LUVIT pointings).  For differences of $|\Delta m| < 0.2$, the difference in the recovery fraction curves and the bias curves are small.  This can be seen by comparing, within a given target, the recovery fraction and bias curves of the lowest and highest stellar density bins for filters with $|\Delta m|<0.2$ in Figure~\ref{fig:completeness_bias_spatial}.

A few general patterns are apparent across the LUVIT pointings.  The NUV observations show no significant impact from crowding on either the limiting magnitudes of the observations or bias in the recovered magnitudes.  However, most LUVIT pointings reach sufficient observed stellar densities in the optical bands to impact the depth of the stellar catalogs in the higher density regions.  A number of the LUVIT pointings also show impacts from stellar crowding in the NIR bands. These impacts are typically larger in F110W than in F160W, which is likely due to the fact that the F110W observations are also typically $\gtrsim 1$~magnitude deeper than the F160W observations. For the targets and filters included in Figure~\ref{fig:completeness_bias_spatial}, the completeness limits are generally larger, and the absolute magnitude of the bias (particularly the 16th- and 84th-percentiles of the distribution) generally smaller, in the lower spatial density bins than in the higher spatial
 density bins.

The two galaxies with the largest spread of $\Delta m$ values are NGC~4163 and UGC~9240.  These two galaxies have a large number of stars passing our quality criteria (Section~\ref{sec:data_redux_catalogs}) in the optical bands combined with some of the highest central stellar surface densities in the LUVIT sample, with the large majority of the stars concentrated within a portion of one HST/ACS chip (Figure~\ref{fig:stellar_spatial_dist}).  In contrast, the WLM-POS2 LUVIT target has the highest number of stars  passing our quality criteria in the optical of any LUVIT pointing, but they are much more evenly spread across the HST/ACS field of view.  The very large $\Delta m_{50\%}$ seen in F555W and F814W in Sextans~A is due to the significant heterogeneity in the depth of the archival observations used in the LUVIT reductions (Section~\ref{sec:notes_select_targets}).   

For a small number of targets, the full-filter region misses areas of higher stellar density which fall outside the full-filter region.  For most of these targets (WLM-POS2, UGC8651, UGC8833, and UGCA292), these are relatively isolated, small areas of higher density.  Relative to these targets, WLM-POS1 has a significantly larger share of area at higher stellar density that falls outside the full-filter region, while the highest stellar density areas in Sextans~A and Sextans~B fall outside the full-filter region.  

Finally, we remind the reader that a few of the LUVIT targets have spatial variation in observed depth for a given archival filter due to the inclusion of multiple pointings in that filter with different depths (Sextans~A, noted above, is an extreme example).  For these targets, some portion, or even the majority, of the differences in the recovery fraction and bias as a function of stellar density may be due to correlations with the observational heterogeneity and not to the impacts of stellar crowding.   

In summary, the impact of blended sources on the photometry in higher (observed) stellar density regions is relatively small for most of the filters in most of the LUVIT stellar catalogs.  However, there are a number of targets where the recovery fraction and bias differ significantly with sky location for some subset of the filters.  For these targets, any analyses that extend to magnitudes at which the recovery fraction is $<1$ should take care to understand the potential impact on their results of the differences in recovery fraction and measurement bias as a function of spatial position. 

\section{Notes on Select Targets}\label{sec:notes_select_targets}

\textit{NGC~3741---} This galaxy is one of the most spatially compact in the LUVIT sample. In order to optimize the spatial distribution of the input ASTs and maximize the number of source density bins used in the AST tests, the ASTs were run only in the chip of the drizzled ACS F475W reference image where the galaxy is located.

\textit{NGC~4163---} This target has both ACS/WFC F475W (archival; Table~\ref{table:archivalpointings}) and WFC3/UVIS F475W (\marthago; Table~\ref{table:martha_primarypointings}) imaging.  The archival ACS/WFC imaging provides a larger footprint, covering more of the optical extent of the galaxy, than the WFC3/UVIS imaging.  The data reduction pipeline does not natively handle images from different instruments with the same filter names, and including them requires special processing.  Utilizing the F475W imaging from both instruments would not increase the total coverage area and the images cannot be combined to increase the depth in F475W. Thus, the LUVIT reductions use only the archival ACS/WFC imaging in order to maximize the area of the galaxy that is covered by the F475W band.  

\textit{SEXTANS-A ---}  The recent SF in Sextans~A is spread over a relatively large spatial extent compared to the FOV of the Hubble instruments.  In addition, Sextans~A has been a target of interest for a large number of HST GO programs with varied scientific goals.  In order to maximize the spatial and spectral coverage of the SF in the LUVIT reductions for Sextans~A, we adopted an inclusive approach to the archival imaging.  Thus, this target has a particularly heterogeneous dataset, which includes both shallow and deep exposures in a large number of cameras and filters.  The archival datasets included in the LUVIT reductions either increased depth or filter coverage, or expanded spatial coverage of regions of recent SF. This target required bespoke changes to the catalog generation steps of the data reduction pipeline, to enable inclusion of images from different instruments with the same filter name. 

The heterogeneity of the datasets results in severe differences in depth as a function of position for some of the archival filters, even within the full-filter region.  A particularly egregious case is that of the WFPC2 F555W and F814W observations (e.g., see the Sextans~A plots in the on-line figure sets Figures~\ref{fig:cmds}, ~\ref{fig:phot_unc}, and \ref{fig:completeness_bias_spatial}, as well as Figure~\ref{fig:crowding_summary}). Because of this, we do not include representative completeness limits for the WFPC2 F555W and F814W observations for Sextans~A in Table~\ref{table:completeness}.  Moreover, we caution that any analysis of the Sextans~A stellar catalogs will need to account for the significantly varying depth, in multiple filters, across the spatial extent of the catalogs. 

\textit{WLM-POS1---} 
As discussed in Section~\ref{sec:obs_strategy_nuv}, the \karriego\ UV observations for WLM-POS1 were shifted slightly with respect to the existing deep archival optical imaging, so that the UV observations overlapped with the detected CO emission in WLM.  As a consequence, shallow WFPC2 optical data in F439W was included in the WLM-POS1 \karriego\ reduction in order to provide photometric measurements in an optical filter for the brightest stars in the region of the CO emission; these images were also subsequently included in the \marthago\ reduction.  
The F439W imaging has only marginal overlap with the rest of the archival WLM-POS1 observations (Figure~\ref{fig:cmds_nfilts}).  Therefore, the F439W imaging footprint was not considered when determining the spatial region from which the CMDs in Figure~\ref{fig:cmds_nfilts} are drawn, and over which the completeness and bias for this target was computed (Section~\ref{sec:data_results_asts}, Table~\ref{table:completeness}).  In other words, what is presented as, and computed from, the ``full-filter region'' for this target is the region where measurements in all filters excluding F439W are possible. The exclusion of F439W from the computation of the full-filter region results in a more representative picture of the completeness and bias for this target for the majority of the filters.  For the completeness and bias analysis for the F439W filter itself, as well as the CMDs that include F439W in Figures~\ref{fig:cmds} and \ref{fig:cmds_nfilts}, we restrict the analysis to th
e true full-filter region (considering all filters including F439W), which is the small sliver of overlap between the adjoining WFPC2 optical footprints seen in the WLM-POS1 plot in Figure~\ref{fig:footprints}.

\section{Selected Science Applications}\label{sec:science_apps}

The LUVIT NUV through NIR photometric catalogs are anticipated to enable a broad range of scientific investigations.  We highlight a few here which focus primarily on the inclusion of the NUV to the dataset; Boyer et al.\ (in prep) will highlight applications focused on the broad- and medium-band NIR observations. 

\subsection{Increased Fidelity of the Recent Star Formation History}
While optical CMD fitting provides a reliable SFH across cosmic time \citep[e.g.,][]{gallart2005, tolstoy2009}, the LUVIT data demonstrate that the NUV CMDs have the potential to better constrain both the younger populations which dominate the UV flux, and the dust properties around the younger stellar populations. We are quantifying the extent to which integrating the NUV CMDs with the optical CMDs enhances the precision and accuracy of recent SFH determinations. Using LUVIT NUV and optical stellar photometry, we can quantify the impact of including NUV data on measuring the SFH by simultaneously modeling the NUV and optical CMDs for the first time (Y.~Choi et al., in prep).

\subsection{Stellar Spectral Energy Distribution Fitting}

A primary motivation for obtaining NUV through NIR imaging in the LUVIT galaxies was to enable SED fitting of resolved stars, thereby constraining both stellar [e.g., age, mass, and metallicity] and line-of-sight dust extinction parameters [dust column ($A_V$), grain size distribution ($R_V$), and the amount of mixing between the Milky Way- and Small Magellanic Cloud-type extinction curves ($f_A$)] in the LUVIT galaxies.  

In addition to enabling the creation of spatially resolved maps of the mean stellar and dust extinction parameters in LUVIT galaxies, such analyses can also be used to infer the intrinsic flux being output by the stars in a galaxy, including at wavelengths not included in the original observations.  For example, \citet{choi2020} used the BEAST to perform SED fits on individual stars in a similarly broad panchromatic dataset in NGC~4214.  The SED fits were used to infer the total ionizing flux being output by the stars in that galaxy, resulting in observational constraints on the spatially resolved, local escape fraction of ionizing photons and yielding a more accurate, and considerably higher than previously found, estimate of the global escape fraction for this high-$z$ analog \citep{choi2020}. 

The NUV-NIR CMD locations and SED fitting has also been used to identify and characterize massive stars in nearby galaxies. For example, \citet{lindberg2024} used BEAST fits to identify over 42 thousand massive star candidates across the PHAT region of M31 \citep{dalcanton2012phat}. In Leo~A, \citet{gull2022} used the LUVIT stellar catalog to identify massive star candidates for spectroscopic observations, prioritizing the brightest stars in the F336W vs.\ (F336W\,--\,F475W) CMD.  Using the BEAST, \citet{gull2022} compared stellar parameters derived from SED fitting of the LUVIT photometry of these massive star candidates with parameters derived by fitting the stellar spectra (specifically, comparing the effective stellar temperature, surface gravity, and stellar luminosity).  They found remarkably good agreement in these parameters for main-sequence (MS) stars with 
satisfactory fits in both methods, indicating that NUV through NIR photometric SED fitting is promising for identifying and characterizing the most massive stars in low-metallicity galaxies too distant in which to obtain resolved stellar spectra. They also found that B-type stars with emission lines stand out as outliers in the SED fits, and are well distinguished from massive MS stars when comparing the F275W vs.\ (F275W\,--\,F475W) CMD to the F475W vs.\ (F475W\,--\,F814W) CMD.

\section{Summary}\label{sec:conclusions}

The LUVIT program combines NUV through NIR HST imaging from \karriego, \marthago, and the HST archive to produce simultaneous multiband, PSF-fit photometric catalogs of the resolved stellar populations for 23 pointings in 22 nearby low-mass, star-forming galaxies.  The LUVIT galaxies range in distance from $\sim 0.75$\,--\,3.9~Mpc, and include some of the lowest mass, lowest metallicity, star-forming galaxies in the Local Volume (Section~\ref{sec:sample}).  

The addition of broadband NUV \karriego\ and broad- and narrow-band NIR (as well as some supplementary optical) \marthago\ observations to the extensive optical observations available in the HST archive provide the missing data needed to address the scientific goals of LUVIT (Section~\ref{sec:intro}).  LUVIT aims to enable quantitative constraints on the time and energy scales driving the cycle of gas and stars in galaxies. This includes enabling direct measurements of both stellar and dust parameters in star-forming regions, mapping the dust, SF, and its evolution over time at scales of tens to hundreds of parsecs, constraining the timescales of the most recent SF in the LUVIT galaxies, and directly observing the youngest stellar sources which contribute the majority of the FUV and NUV flux of the galaxies. It also includes studying the evolved massive star populations, a dominant source of the optical and NIR flux in star-forming galaxies, in order 
to provide observational constraints of the dependence on metallicity of the physical processes that drive evolved star models (e.g., mass loss, dredge up, convection overshoot, rotation).  These stellar evolution models in turn strongly influence the quantities derived from unresolved light in more distant galaxies (e.g., SFR, total stellar mass, and metallicity).  

The observing strategy (Section~\ref{sec:survey_design}) and data reduction techniques (Section~\ref{sec:data_redux}) were optimized for studying the properties of the individual stars in the LUVIT galaxies.  The resulting stellar catalogs (Section~\ref{sec:data}) show an interesting diversity of morphology in the CMDs, which can be further quantified in the future via CMD-based SFH analyses.  The NUV observations are in nearly all cases the shallowest of the available bands.  Thus, the stellar populations which are measured in six or more broad filters (two NUV, at least two optical, and two NIR) are generally the brightest (most massive) MS stars and blue helium-burning stars in the LUVIT galaxies. 

Artificial star tests were used to quantify the photometric quality of the stellar catalogs (Section~\ref{sec:data_redux_asts}).  The analysis of the ASTs presented here provides a representative view of the quality of the data over the spatial regions in each LUVIT pointing in which data exists in all available filters.  Given the complexity of the LUVIT datasets, both in the heterogeneity of the archival observations and in the significant spread in observed stellar surface density within many of the LUVIT pointings, many scientific analyses will need to utilize the full results of the ASTs rather than representative average values.  A few targets with particularly heterogeneous datasets were discussed in Section~\ref{sec:notes_select_targets}.

This paper focused on the broadband LUVIT observations.  A future companion survey paper will discuss the narrow-band NIR data obtained as part of \marthago, and will provide relevant details for the future LUVIT data release and a full dataset DOI (Boyer et al., in prep).  A number of  specific science investigations are currently being undertaken with the LUVIT data (Section~\ref{sec:science_apps}). While these include ones originally envisioned for the LUVIT program, the LUVIT dataset is also inspiring uses of the panchromatic imaging which were not originally planned \citep{gull2022}, highlighting the power of high-resolution panchromatic imaging for the study of resolved stellar populations in nearby galaxies.

\acknowledgements
Support for this work was provided by NASA through grants \#\karriego, \#\marthago, and \#GO-16292 from the Space Telescope Science Institute.  This research is based on observations made with the NASA/ESA Hubble Space Telescope obtained from the Space Telescope Science Institute, which is operated by the Association of Universities for Research in Astronomy, Inc., under NASA contract NAS 5–26555. These observations are associated with programs \karriego\, \marthago, and GO-16292.

This work was based on observations made with the NASA/ESA Hubble Space Telescope, and obtained from the Hubble Legacy Archive, which is a collaboration between the Space Telescope Science Institute (STScI/NASA), the Space Telescope European Coordinating Facility (ST-ECF/ESAC/ESA) and the Canadian Astronomy Data Centre (CADC/NRC/CSA).

\facilities{HST(ACS/WFC), HST(WFC3/UVIS), HST(WFC3/IR), HST(WFPC2)} 

Software: \texttt{Astropy} \citep{AstropyCollaboration2013,AstropyCollaboration2018,AstropyCollaboration2022}, 
\texttt{BEAST} \citep{gordon2016},
\texttt{DOLPHOT} \citep{hstphot, dolphin2016}, 
\texttt{Drizzlepac} (\citealt{STSCI2012drizzle,Hack2013drizzle,Avila2015drizzle}), 
\texttt{Matplotlib} \citep{hunter2007}, 
\texttt{NumPy} \citep{vanderwalt2011, harris2020}.
\texttt{SciPy} \citep{Virtanen2020}.

\bibliography{new.ms}

\newcolumntype{L}[1]{>{\raggedright\arraybackslash}p{#1}}
\newcolumntype{R}[1]{>{\raggedleft\arraybackslash}p{#1}}

\begin{longrotatetable}
\movetabledown=6mm
 
\appendix

\section{Comparison of the \marthago\ and \karriego\ Reductions}\label{app:stack_comparison}

The LUVIT data reduction, photometric measurements, and resulting stellar catalogs presented in this paper incorporate the \karriego, \marthago, and archival datasets, with the exception of the three LUVIT targets not included in \marthago\ (DDO6, M81-DWARF-A, and UGC8760).  These three targets were reduced directly following their \karriego\ observations, as were all LUVIT targets.  In addition, select science applications may benefit from utilizing the initial \karriego\ plus archival reductions in some targets, due to the reference image footprint from the \karriego\ reductions covering slightly more of the star-forming region of the galaxy.  Thus, this section discusses the differences between the initial (\karriego\ and archival) and primary (\karriego, \marthago, and archival) LUVIT reductions.

All reductions were run using images downloaded from the Hubble archive shortly after completion of the observations for each of the \karriego\ or \marthago\ targets.  They thus incorporate the best-known calibrations and astrometric solutions at the time of download.\footnote{All images for a given target, including archival observations and those from \karriego, were re-downloaded from MAST upon completion of each \marthago\ observation.} 
While no changes were made to our reduction pipeline between the \karriego\ and \marthago\ reductions,  
there were two updates to the Hubble calibration pipelines of potential significance for the LUVIT reductions. 
First, Gaia sources were added as astrometric reference stars\footnote{https://hst-docs.stsci.edu/drizzpac/chapter-4-astrometric-information-in-the-header/4-5-absolute-astrometry} , improving the accuracy and precision of the WCS solutions in data downloaded from MAST. While this did not significantly impact the precision of registering individual images to the reference image when running DOLPHOT, 
in some cases it did improve alignment by enabling the use of a different reference image, as discussed below.  
Second, data products produced after 2020 October (including all observations downloaded for the \marthago\ reductions) incorporated an updated, time-dependent, calibration for the WFC3 detectors \citep{calamida2022}, included in the metadata of CALACs products.  However, DOLPHOT does not support multiple zero-points, and thus these are not incorporated into the stellar catalogs. 
For the datasets used here, it is a $\sim 0.1$\% per year effect, which is well below other sources of error (such as PSF/focus variations) in the LUVIT reductions. 

The largest impact on the quality of the stellar catalogs between the initial reductions of the \karriego\ and archival observations, and the primary reductions incorporating the \marthago\ observations, was the ability to use F475W as the reference image for alignment of the individual images for the simultaneous multiband photometry.  The \marthago\ observations added F475W observations to all targets lacking this filter, and the improved WCS solutions from the addition of Gaia sources as astrometric reference stars enabled the use of existing archival F475W data for the reference image for some targets in which another filter was used as the reference image in the initial reductions (e.g.,  DDO210).  This resulted in successful multiband alignment using the drizzled F475W image as the reference image for all targets included in \marthago.  
In general, the use of F475W for construction of the reference image both improved the initial relative astrometric alignment of the individual images and reduced the significance of the residuals of the registration of the individual chips and images with the reference image in Dolphot.  

The footprints of the F475W reference images generally include either all or the vast majority of the WFC3/UVIS NUV footprint.  However, for a small number of targets the loss of spatial coverage of the NUV when the F475W reference image is used 
(Section~\ref{sec:data_redux_image_processing})
may be detrimental to analyses focused on NUV observations and sources.  These targets are 
UGC8508, LeoA, DDO210, and UGC7577.  For these targets, we will also release photometric catalogs (Boyer et al., in prep) derived from the original \karriego\ plus archival reductions, 
which utilize reference images that cover a larger portion of the field of view of the UV observations. 
Table~\ref{table:uvstack_refimage} lists the camera and filter of the observations from which the combined, drizzled reference image was derived for these six targets, as well as the pipeline name of these targets in the original \karriego\ plus archival reductions. Table~\ref{table:uvstack_refimage} also includes the camera and filter used to create the reference image for the three targets not included in \marthago. 

\end{document}